\documentclass{article}
\usepackage[utf8]{inputenc}
\usepackage{latexsym,amsmath,amscd,amssymb,framed,graphics,mathrsfs}
\usepackage[colorlinks]{hyperref}
\usepackage[margin=1cm]{geometry}
\usepackage{frame}
\usepackage{siunitx}

\newtheorem{theorem}{Theorem}[section]

\newtheorem{remark}[theorem]{Remark}

\newcommand{\pp}[2]{\frac{\partial #1}{\partial #2}}

\newcommand{\DD}[1]{\frac{D \!#1}{D t}}

\title{Thermodynamically consistent versions of approximations used in modelling moist air}
\author{Christopher Eldred, Mark Taylor, Oksana Guba}

\begin{document}

\maketitle
\tableofcontents

\abstract{Some existing approaches to modelling the thermodynamics of moist air make approximations that break \textit{thermodynamic consistency}, such that the resulting thermodynamics do not obey the 1st and 2nd laws or have other inconsistencies. Recently, an approach to avoid such inconsistency has been suggested: the use of \textit{thermodynamic potentials} in terms of their \textit{natural variables}, from which all thermodynamic quantities and relationships (equations of state) are derived. In this paper, we develop this approach for \textit{unapproximated} moist air thermodynamics and two widely used approximations: the constant $\kappa$ approximation and the dry heat capacities approximation. The (consistent) constant $\kappa$ approximation is particularly attractive because it leads to, with the appropriate choice of thermodynamic variable, adiabatic dynamics that depend only on total mass and are independent of the breakdown between water forms. Additionally, a wide variety of material from different sources in the literature on thermodynamics in atmospheric modelling is brought together. It is hoped that this paper provides a comprehensive reference for the use of thermodynamic potentials in atmospheric modelling, especially for the three systems considered here.}

\section{Introduction}

When considering moist air, it is easy to introduce approximations that break \textit{thermodynamic consistency}, such that the resulting thermodynamics do not obey the 1st and 2nd laws or have other inconsistencies. For example, a common approach is to use \textit{unapproximated} thermodynamics for moist air but define $\theta_v$ through $p \alpha = \kappa_d \Pi_d \theta_v$ (with $\Pi_d = C_{pd} (p / p_r)^{\kappa_d}$ and $\kappa_d = R_d / C_{pd}$) and treat it as an \textit{advected quantity}\footnote{An \textit{advected quantity} $a$ obeys $\DD{a} = 0$ for reversible dynamics.}. This is known as the constant $\kappa$ approximation, and applied in this way it is thermodynamically inconsistent: the treatment of $\theta_v$ as an advected quantity leads to equations that no longer conserve the total energy, since there are missing water-related terms that should appear on the right hand side of the equation (i.e. $\theta_v$ is not really an advected quantity).

\begin{remark}
Although we have referred here to \textit{unapproximated} thermodynamics for moist air, such a thing does not actually exist. All thermodynamic potentials are \textit{empirical}: they come from either experiment or derivation from a more fundamental underlying theory such as molecular dynamics or statistical mechanics. For example, in this work we assume that heat capacities are temperature-independent, condensates occupy no volume and that each phase of water is an unrelated thermodynamic substance (i.e. there is no equilibrium between water species and Gibbs phase rule does not apply). This highlights the fundamental difference between \textit{inconsistency error} (which is avoidable) and \textit{approximation error} (which is unavoidable).
\end{remark}

A general approach to avoid such inconsistency is through the use of \textit{thermodynamic potentials}, from which all thermodynamic quantities and relationships can be derived. This approach was advocated in \cite{Thuburn2017,Thuburn2017a,Staniforth2019}, although complete sets of thermodynamic potentials in terms of their \textit{natural variables} are not presented in those works. An important usage for potentials is a more rigorous treatment of energetics within a modelling system, as discussed in \cite{Lauritzen2020}. Internal energy in terms of its natural variables is presented in \cite{Bowen2022a,Bowen2022b}. In this paper, we consider two widely used approximations: the constant $\kappa$ approximation and the dry heat capacities approximation. These approximations are used in many existing atmospheric models, often inconsistently. In fact, it is usually difficult to determine exactly what thermodynamic potentials are used for a given model or model component; or even if there is a single set of thermodynamics. If there is documentation of the thermodynamics, it usually consists of some equations of state\footnote{Following the usual atmospheric dynamics literature terminology, which differs from the terminology in the thermodynamics literature.} such as $p \alpha = R^*T$ and $\theta = T (\frac{p}{p_r})^\frac{R^*}{C_p^*}$, which are possibly independently approximated. However, as demonstrated in this paper, equations of this type do not completely specify the thermodynamics. Additionally, independently approximating these expressions can lead to inconsistency. Some consequences of this inconsistency are explored in Section \ref{inconsistency}. By instead starting with thermodynamic potentials and introducing the relevant approximations directly into these, inconsistency can be avoided. We refer to the consistent systems derived from the approximate potentials as the \textit{constant $\kappa$ system} and the \textit{dry heat capacities system}.  In addition to consistency, the constant $\kappa$ system also simplifies the dependence on the various  water forms (vapor, liquid and ice) so that with an appropriate choice of thermodynamic variable (such as the virtual potential temperature $\theta_v$ discussed below) the adiabatic dynamics depend only on the total mass and the water forms decouple from the rest of the dynamics.  This is attractive for numerical modelling, since then water forms can be advanced independently from the dynamics with a larger timestep and/or different numerics.

There are some modelling systems (as of the publication of this paper) where the thermodynamics are explicitly articulated and a thermodynamic system can be identified. These are:
\begin{itemize}
    \item CAM-SE/E3SMv1 \cite{Lauritzen2019}: uses the dry heat capacities system
    \item E3SMv2 (specifically the HOMME-NH \cite{Taylor2020} dynamical core): uses the constant $\kappa$ system
    \item CAM-SE-CSLAM \cite{Lauritzen2019}: uses the unapproximated system 
\end{itemize}
It is hoped that this paper will encourage more groups to explicitly articulate the thermodynamics utilized in their models, and stimulate future research into new consistent thermodynamics with different assumptions.

The main material in this paper is complete sets\footnote{The four commonly used thermodynamic potentials: internal energy, enthalpy, Gibbs function and Helmholtz free energy.} of thermodynamic potentials in their natural variables for the \textit{unapproximated}, constant $\kappa$ and dry heat capacities systems.  This work also brings together a lot of material that is scattered around in various sources in the literature, and attempts to provide a comprehensive reference for the use of thermodynamic potentials in atmospheric modelling. It builds on the Gibbs functions introduced in \cite{Thuburn2017,Thuburn2017a} and the internal energies introduced in \cite{Lauritzen2019,Staniforth2019}, specifically in making the same fundamental assumptions and obtaining the same thermodynamic potentials up to certain linear functions which just shift the zeros of entropy and chemical potentials, as discussed in Appendix \ref{deriv-unapprox}. However, \cite{Thuburn2017,Thuburn2017a,Lauritzen2019,Staniforth2019} do not present all thermodynamic potentials and/or do not give them in terms of their natural variables, both of which are done here.

The remainder of this paper is structured as follows: Section \ref{review-equilbrium-thermo} presents a review of equilbrium thermodynamics in the general case, Section \ref{unapprox-moist-air} provides the thermodynamic potentials and related quantities for \textit{unapproximated} (in the sense discussed above) moist air, Section \ref{approx-moist-air} provides the thermodynamic potentials and related quantities for the constant $\kappa$ and dry heat capacities systems and finally Section \ref{conclusions} gives some conclusions. Appendix \ref{deriv-unapprox} derives the \textit{unapproximated} potentials found in Section \ref{unapprox-moist-air}, Appendix \ref{potential-virtual-quantities} discusses potential and virtual quantities, Appendix \ref{common-thermo-quantities} gives common thermodynamic quantities for all three systems, Appendix \ref{latent-heats} gives latent heats for all three systems and Appendix \ref{chemical-potentials} gives chemical potentials for all three systems. It is hoped that the material in Sections \ref{review-equilbrium-thermo} - \ref{approx-moist-air} and the Appendices provides a comprehensive reference for the use of thermodynamic potentials in atmospheric modelling, especially for the three systems considered. 

\section{Review of Equilibrium Thermodynamics}
\label{review-equilbrium-thermo}
Consider a multispecies, multiphase fluid composed of $N$ components. By components here we refer to constituents with distinct thermodynamic behavior. In addition to separate substances, this can also include different phases of the same species (ex. water vapor and liquid water) and/or different allotropes of the same species and phase (ex. ortho and para forms of hydrogen). In writing the thermodynamics below, we do not assume any equilibrium between different phases of the same species or chemical components undergoing reactions; and instead treat each phase as an independent thermodynamic substance. In particular, this means that each phase has it's own independent density, rather than a total density for that species with proportioning between phases done according to some sort of equilibrium hypothesis or phase rule. This assumption fits with the commonly used splitting in atmospheric modelling between (adiabatic) dynamics and physics. We will also assume that all components of the fluid are at the same temperature $T$\footnote{Relaxing the single temperature approximation is possible, but leads to significant increases in complexity. For an example of this in moist atmospheric dynamics, see \cite{Bannon2002}.}.

A key assumption made in modelling this fluid is that local thermodynamic equilibrium (LTE) holds, in the sense that thermal relaxation times are sufficiently short compared to other dynamical time scales and therefore large scale thermodynamic quantities can be meaningfully defined; for example temperature and entropy. LTE implies that the thermodynamics of this fluid can be described using equilibrium thermodynamics, which are reviewed below. The assumption of LTE is almost universal in atmospheric modelling, especially for the troposphere and stratosphere. More information on equilibrium thermodynamics can be found in standard textbooks on the subject, such as \cite{Zdunkowski2004}.

\subsection{Thermodynamic Potentials}
Equilibrium thermodynamics tells us that the thermodynamic behaviour of this fluid is completely determined by a \textit{thermodynamic potential} written in terms of its \textit{natural (state) variables}. In the thermodynamics literature the specification of a thermodynamic potential in terms of its natural variables is referred to as the \textit{equation of state}, while a somewhat different usage occurs in the atmospheric dynamics literature (see below). In the case of the fluid described above, the state variables are one choice from the conjugate pair (volume $V$, pressure $p$), one choice from the conjugate pair (entropy $S$, temperature $T$) and one choice for each component from the conjugate pair (component mass $M_n$, component chemical potentials $\mu_n$)\footnote{This presentation is slightly different than the standard one, which would work in terms of component molar quantities $N_n$ or component number of particles $(N_{i})_n$, which are related to mass by $M_n = m_n N_n = m_n N_a (N_{i})_n$ through the molar mass $m_n$ and the Avogadro constant $N_a$.}. Note that one member of each pair is an \textit{extensive quantity}, while the other is an \textit{intensive quantity}. In atmospheric dynamics, four thermodynamic potentials are commonly used: the \textit{internal energy} $U(V,S,M_n)$, the \textit{enthalpy} $H(p,S,M_n)$, the \textit{Gibbs free energy} $G(p,T,M_n)$ and the \textit{Helmholtz free energy} $F(V,T,M_n)$. These are related through
\begin{eqnarray}
H &=& U + p V\\
G &=& U + p V - S T\\
F &=& U - S T
\end{eqnarray}
which are known as \textit{Legendre transforms}.

Associated with each of these thermodynamic potentials are conjugate variables:
\begin{eqnarray}
\label{u-conj}
p(V,S,M_n) := -\pp{U}{V} \quad\quad\quad T(V,S,M_n) := \pp{U}{S} \quad\quad\quad \mu_n(V,S,M_n) := \pp{U}{M_n} \\
V(p,S,M_n) := \pp{H}{p} \quad\quad\quad T(p,S,M_n) := \pp{H}{S} \quad\quad\quad \mu_n(p,S,M_n) := \pp{H}{M_n} \\
V(p,T,M_n) := \pp{G}{p} \quad\quad\quad S(p,T,M_n) := -\pp{G}{T} \quad\quad\quad \mu_n(p,T,M_n) := \pp{G}{M_n} \\
\label{f-conj}
p(V,T,M_n) := -\pp{F}{V} \quad\quad\quad S(V,T,M_n) := -\pp{F}{T} \quad\quad\quad \mu_n(V,T,M_n) := \pp{F}{M_n}
\end{eqnarray}
Note that a given conjugate variable has multiple expressions: for example we have $p(V,T,M_n)$ and $p(V,S,M_n)$. These are not in fact different quantities, they are just alternative ways of expressing the same quantity in terms of different variables.

Using (\ref{u-conj}) - (\ref{f-conj}) plus the total differential, the \textit{fundamental thermodynamic identities} for $U(V,S,M_n)$, $H(p,S,M_n)$, $G(p,T,M_n)$ and $F(V,T,M_n)$ are:
\begin{eqnarray}
\label{u-fund-thermo}
dU &=& -p dV + T dS + \sum_n \mu_n dM_n\\
\label{h-fund-thermo}
dH &=& V dp + T dS + \sum_n \mu_n dM_n\\
\label{g-fund-thermo}
dG &=& V dp - S dT + \sum_n \mu_n dM_n\\
\label{f-fund-thermo}
dF &=& -p dV - S dT + \sum_n \mu_n dM_n
\end{eqnarray}
Utilizing (\ref{u-fund-thermo}) - (\ref{f-fund-thermo}) along with Euler's homogeneous function theorem gives expressions for the thermodynamic potentials (up to a constant) as
\begin{eqnarray}
\label{u-expr}
U &=& -p V + T S + \sum_n \mu_n M_n \\
H &=& T S + \sum_n \mu_n M_n \\
G &=& \sum_n \mu_n M_n \\
\label{f-expr}
F &=& -p V + \sum_n \mu_n M_n
\end{eqnarray}
In particular this shows that the chemical potentials $\mu_n$ are just the partial Gibbs free energies. Finally, combining (\ref{u-fund-thermo}) - (\ref{f-fund-thermo}) and (\ref{u-expr}) - (\ref{f-expr}) gives the \textit{Gibbs-Duhem relationship}
\begin{equation}
    0 = - V dp + S dT + \sum_n M_n d\mu_n 
\end{equation}
This shows that the intensive quantities are not all independent, and leads to the \textit{Gibbs phase rule}.

%\todo{WHAT ARE NATURAL VARIABLES? WHAT ARE CONJUGATE PAIRS? Thermo potentials are types of energy?}

\subsection{Specific Thermodynamic Potentials}
In atmospheric fluid dynamics, it is common to work in terms of \textit{specific} quantities (quantities per unit mass) rather than extensive quantities. In doing so, introduce \textit{specific volume} $\alpha = \frac{V}{M}$, the \textit{specific entropy} $\eta = \frac{S}{M}$ and the \textit{specific component concentration} $q_n = \frac{M_n}{M}$, where $M = \sum_n M_n$ is total mass. Finally, consider the \textit{specific internal energy} $u(\alpha,\eta,q_n)$, the \textit{specific enthalpy} $h(p,\eta,q_n)$, the \textit{specific Gibbs free energy} $g(p,T,q_n)$ and the \textit{specific Helmholtz free energy} $f(\alpha,T,q_n)$. These can be related to $U(V,S,M_n)$, $H(p,S,M_n)$, $G(p,T,M_n)$ and $F(V,T,M_n)$ through
\begin{equation}
\label{specific-to-full-potentials}
    U = M u(\frac{V}{M}, \frac{S}{M}, \frac{M_n}{M}) \quad\quad H = M h(p, \frac{S}{M}, \frac{M_n}{M}) \quad\quad G = M g(p, T, \frac{M_n}{M}) \quad\quad F = M f(\frac{V}{M}, T, \frac{M_n}{M})
\end{equation}
\begin{remark}
Given specific variables $\alpha$, $\eta$ and $q_n$, it is only possible to determine $U$, $H$, $G$ and $F$ up to a constant multiplier, since $M$ cannot be obtained from these purely specific quantities. However, the partial derivatives of the thermodynamic potentials are what determines the thermodynamics of a system, and therefore this is not an impediment to use of specific thermodynamic potentials $u$, $h$, $g$ and $f$ instead of $U$, $H$, $G$ and $F$ .
\end{remark}

Exactly as before, the specific thermodynamic potentials are related through Legendre transforms as
\begin{eqnarray}
h &=& u + p \alpha\\
g &=& u + p \alpha - \eta T\\
f &=& u - \eta T
\end{eqnarray}
Using (\ref{u-conj}) - (\ref{f-conj}), (\ref{specific-to-full-potentials}) and the chain rule , it is not difficult to show that 
\begin{eqnarray}
\label{U-conj}
p(\alpha,\eta,q_n) = -\pp{u}{\alpha} \quad\quad\quad T(\alpha,\eta,q_n) = \pp{u}{\eta}\\
\label{H-conj}
\alpha(p,\eta,q_n) = \pp{h}{p} \quad\quad\quad T(p,\eta,q_n) = \pp{h}{\eta}\\
\label{G-conj}
\alpha(p,T,q_n) = \pp{g}{p} \quad\quad\quad \eta(p,T,q_n) = -\pp{g}{T}\\
\label{F-conj}
p(\alpha,T,q_n) = -\pp{f}{\alpha} \quad\quad\quad \eta(\alpha,T,q_n) = -\pp{f}{T}
\end{eqnarray}
However, the dependence on $M_n$ (through $M$) leads to somewhat complicated expressions for $\mu_n$ in terms of $\pp{x}{q_n}$, and that $\mu_n \neq \pp{x}{q_n}$, for any of $x \in (u,h,g,f)$. These expressions are deferred to Appendix \ref{chemical-potentials}.

\begin{remark}
In the atmospheric dynamics literature, (\ref{U-conj}) - (\ref{F-conj}) are themselves often referred to as the \textit{equation of state}. For example, the expression of $p$ in terms of $\alpha$, $T$ and $q_n$ for a single component perfect ideal gas is $p \alpha = R T$, usually given the name equation of state. However, this expression itself does not contain sufficient information to completely determine the thermodynamics: only the thermodynamic potentials written in terms of their natural variables do. As an example, both the constant $\kappa$ and unapproximated fluids from Section \ref{unapprox-moist-air} and Section \ref{constant-kappa} have $p \alpha = R^\star T$ despite being distinct thermodynamic systems with different behaviour. Additionally, just the potentials themselves are not sufficient, for example the internal energy for a single component perfect ideal gas can be written as $u = C_v T$, but this is not in terms of natural variables and therefore does not have sufficient information to completely determine the thermodynamics.
\end{remark}

Using (\ref{u-conj}) - (\ref{f-conj}) plus (\ref{specific-to-full-potentials}), analogues of the fundamental thermodynamic identities for $u(\alpha,\eta,q_n)$, $h(p,\eta,q_n)$, $g(p,T,q_n)$ and $f(\alpha,T,q_n)$ are:
\begin{eqnarray}
\label{U-fund-thermo}
du &=& -p d\alpha + T d\eta + \sum_n \mu_n dq_n\\
\label{H-fund-thermo}
dh &=& \alpha dp + T d\eta + \sum_n \mu_n dq_n\\
\label{G-fund-thermo}
dg &=& \alpha dp - \eta dT + \sum_n \mu_n dq_n\\
\label{F-fund-thermo}
df &=& -p d\alpha - \eta dT + \sum_n \mu_n dq_n
\end{eqnarray}
Combining (\ref{u-expr}) - (\ref{f-expr}) and (\ref{specific-to-full-potentials}) gives expressions for the specific thermodynamic potentials (up to a constant) as
\begin{eqnarray}
\label{U-expr}
u &=& -p \alpha + T \eta + \sum_n \mu_n q_n \\
h &=& T \eta + \sum_n \mu_n q_n \\
g &=& \sum_n \mu_n q_n \\
\label{F-expr}
f &=& -p \alpha + \sum_n \mu_n q_n
\end{eqnarray}
As before, the chemical potentials $\mu_n$ are just the partial Gibbs free energies. Combining (\ref{U-fund-thermo}) - (\ref{F-fund-thermo}) and (\ref{U-expr}) - (\ref{F-expr}) gives the specific form of the Gibbs-Duhem relationship
\begin{equation}
    0 = - \alpha dp + \eta dT + \sum_n q_n d\mu_n 
\end{equation}
again showing not all the intensive quantities are independent.

\subsection{Entropic Variables}

Instead of using specific entropy $\eta$, it is also possible to use an arbitrary (invertible) function of specific entropy and concentrations: a specific entropic variable $\chi = \chi(\eta, q_n)$. Entropic variables such as potential temperature and potential enthalpy are widely used in atmospheric and oceanic dynamics for two main reasons: in the case of reversible dynamics they remain advected quantities (since specific entropy and concentrations are advected quantities in this case); and for certain thermodynamic potentials a careful choice of entropic variable gives much simpler expressions than specific entropy. Some examples of this are provided in Sections \ref{unapprox-moist-air} and \ref{approx-moist-air}.

Introduce the extensive entropic variable $\Xi = M \chi$, and note that $S$ can be written in terms of $\Xi$ and $M_n$ (by inverting $\chi(\eta,q_n)$ to obtain $\eta(\chi,q_n)$) as
\begin{equation}
\label{S-entropic}
    S = M \eta(\frac{\Xi}{M}, \frac{M_n}{M})
\end{equation}
Now consider $U$ and $H$ as functions of $\Xi$ instead of $S$, denoted with a $\prime$ in this section and Appendix E only\footnote{We rely on context in the remaining sections to determine if we have functions of entropy or an entropic variable.}:
\begin{equation}
\label{full-to-full-entropic}
U^\prime(V,\Xi,M_n) = U(V,M \eta(\frac{\Xi}{M}, \frac{M_n}{M}), M_n)\quad\quad\quad
H^\prime(p,\Xi,M_n) = H(p,M \eta(\frac{\Xi}{M}, \frac{M_n}{M}), M_n)
\end{equation}
Doing so and using the chain rule on partial derivatives yields
\begin{eqnarray}
\pp{U^\prime}{V} = \pp{U}{V} = -p \quad\quad\quad \pp{U^\prime}{\Xi} = \pp{U}{S} \pp{\eta}{\Xi} = T \pp{\eta}{\Xi} = \lambda \\
\pp{H^\prime}{p} = \pp{H}{p} = \alpha \quad\quad\quad \pp{H^\prime}{\Xi} = \pp{H}{S} \pp{\eta}{\Xi} = T \pp{\eta}{\Xi} = \lambda
\end{eqnarray}
where $\lambda = T \pp{\eta}{\Xi}$ is the \textit{generalized temperature}, along with complicated expressions for \textit{generalized chemical potential} $\xi_n = \pp{U^\prime}{M_n} = \pp{H^\prime}{M_n}$ given in Appendix \ref{chemical-potentials}. The fundamental thermodynamic identities for internal energy $U^\prime(V,\Xi,M_n)$ and enthalpy $H^\prime(p,\Xi,M_n)$ can therefore be written as
\begin{eqnarray}
dU^\prime &=& - p d V + \lambda d\Xi + \sum_n \xi_n dM_n\\
dH^\prime &=& V dp + \lambda d\Xi + \sum_n \xi_n dM_n
\end{eqnarray}
In other words, we have replaced the conjugate pairs ($S$, $T$) and ($M_n$, $\mu_n$) with ($\Xi$, $\lambda$) and ($M_n$, $\xi_n$). Finally, we have expressions for $U^\prime$ and $H^\prime$ (up to a constant) as
\begin{eqnarray}
U^\prime &=& -p V + \lambda \Xi + \sum_n \xi_n M_n \\
H^\prime &=& \lambda \Xi  + \sum_n \xi_n M_n
\end{eqnarray}
and a Gibbs-Duhem type relationship
\begin{equation}
    0 = - V dp + \Xi d\lambda + \sum_n M_n d\xi_n 
\end{equation}

Using specific thermodynamic potentials $u(\alpha,\eta,q_n) = u(\alpha, \eta(\chi,q_n),q_n) = u^\prime(\alpha, \chi, q_n)$ and $h(p,\eta,q_n) = h(p, \eta(\chi,q_n),q_n) = h^\prime(p, \chi, q_n)$ , we have the expressions
\begin{equation}
\label{full-entropic-to-specific-entropic}
U^\prime(V,\Xi,M_n) = M u^\prime(\frac{V}{M}, \frac{\Xi}{M}, \frac{M_n}{M})\quad\quad\quad
H^\prime(p,\Xi,M_n) = M h^\prime(p, \frac{\Xi}{M}, \frac{M_n}{M})
\end{equation}
which give
\begin{eqnarray}
p(\alpha,\chi,q_n) &=&  \pp{u^\prime}{\alpha}(\alpha,\chi,q_n)\\
\lambda(\alpha,\chi,q_n) &=&  \pp{u^\prime}{\chi}(\alpha,\chi,q_n) = T(\alpha,\eta(\chi,q_n),q_n)  \eta_\chi(\chi,q_n) \\
\alpha(p,\chi,q_n) &=&  \pp{h^\prime}{p}(p,\chi,q_n)\\
\lambda(p,\chi,q_n) &=&  \pp{h^\prime}{\chi}(p,\chi,q_n) = T(p,\eta(\chi,q_n),q_n)  \eta_\chi(\chi,q_n)
\end{eqnarray}
for the conjugate variables. However, as is the case for $\mu_n$, $\xi_n \neq \pp{x^\prime}{q_n}$ for $x^\prime \in (u^\prime,h^\prime)$. Instead, a complicated relationship between $\xi_n$ and $\pp{x^\prime}{q_n}$ holds, discussed in Appendix \ref{chemical-potentials}. The fundamental thermodynamic identities can be written as
\begin{eqnarray}
du^\prime &=& - p d\alpha + \lambda d\chi + \sum_n \xi_n dq_n\\
dh^\prime &=& \alpha dp + \lambda d\chi + \sum_n \xi_n dq_n
\end{eqnarray}
Finally, we have expressions for $u^\prime$ and $h^\prime$ (up to a constant) as
\begin{eqnarray}
u^\prime&=& -p \alpha + \lambda \chi + \sum_n \xi_n q_n \\
h^\prime &=& \lambda \chi + \sum_n \xi_n q_n
\end{eqnarray}
and a specific Gibbs-Duhem type relationship
\begin{equation}
    0 = - \alpha dp + \chi d\lambda + \sum_n q_n d\xi_n 
\end{equation}

\section{\textit{Unapproximated} Thermodynamics of Moist Air}

\label{unapprox-moist-air}
We will now specialize the general development of equilibrium thermodynamics in Section \ref{review-equilbrium-thermo} to the case of moist, cloudy air: a mixture of dry air and the three phases of water: water vapor, (cloud) liquid water and (cloud) ice. The concentrations of these four components are denoted with $q_d$, $q_v$, $q_l$ and $q_i$, respectively. The gaseous components $q_d$ and $q_v$ are assumed to be perfect ideal gases with temperature-independent) (i.e. constant) heat capacities at constant volume $C_{vd}$ and $C_{vv}$ and at constant pressure $C_{pd}$ and $C_{pv}$. The condensed components $q_l$ and $q_i$ are assumed to be incompressible with temperature-independent heat capacities $C_l$ and $C_i$ and to appear as pure substances, and we neglect the volume occupied by the condensates\footnote{Condensate volume can be incorporated without too much additional effort, as done in \cite{Thuburn2017,Staniforth2019,Pelkowski2011}. A more sophisticated treatment of condensates that takes into account the behaviour of droplet and other hydrometeor populations is well beyond the scope of this article, and where the assumption of LTE begins to break down.}. Additionally, we do not assume any equilibrium between phases, so we can capture super saturation and other out of equilibrium situations. 

%Conservation of mass for dry air and water vapor gives $\sum_n \rho_n = \rho$, and therefore $\sum_n q_n = 1$. This latter point is quite important, since as discussed above it means that not all the $q_n$'s are independent. Therefore, in what follows we will let $q_d = 1 - q_v - q_l - q_i$ be the concentration that depends on the others, but keep the symbol $q_d$ since it makes the resulting expressions more symmetric. This will primarily affect the chemical potentials founds in Appendix \ref{chemical-potentials}.
%We also ignore the possibility of precipitation, and of sublimation or deposition (i.e. no direct gas to solid exchanges).

\begin{remark}
Following \cite{Thuburn2017,Staniforth2019}, we could instead predict only total water $q_w = q_v + q_l + q_i$ (i.e. write $u = u(\alpha, \eta, q_d, q_w)$, etc.) and determine $q_v$, $q_l$ and $q_i$ from $q_w$ based on some sort of equilibrium assumption given $p$ and $T$ (or other conjugate pairs). However, this approach leads to certain thermodynamic potentials becoming discontinuous across phase boundaries such as ($p=1000mb$, $T=273.15K$) \cite{Velasco2007}. This can make the definition of conjugate variables variables such as $p = - \pp{u}{\alpha}$ problematic. Additionally, it very unclear how to treat situations such as mixed phase clouds where all three phases occur simultaneously with this approach. One possible approach is the use of a generic "condensed water" substance, as done in \cite{Ooyama1990,Ooyama2001}.
%ie T>0C, have $q_i = 0$ and some mixture of $q_v$ and $q_l$ if enough total water is present, otherwise $q_l =0$; T<0C have $q_l = 0$ and some mixture of $q_v$ and $q_i$ if enough total water is present, otherwise $q_i =0$
\end{remark}

\begin{remark}
Although we refer to the fluid above as \textit{unapproximated} moist air, there are in fact several approximations we have made, such as ideal gas behaviour for dry air and water vapor, temperature-independent heat capacities, zero-volume condensates and that condensates occur in a pure form (without any mixture of types). This is an example of the remark made in the introduction that all thermodynamic potentials are approximate (there are always approximation errors), but it is at least possible to avoid consistency errors through using a single (set of) thermodynamic potentials to derive all thermodynamic relationships.
\end{remark}

Under these assumptions the full expressions for the various specific thermodynamic potentials are
\begin{eqnarray}
\label{U-unapprox}
u(\alpha,\eta,q_d,q_v,q_l,q_i) &=& C_v^* T_r \exp (\frac{\eta - \eta_r}{C_v^*}) (\frac{\alpha}{q_d \alpha_{rd}})^{-\frac{q_d R_d}{C_v^*}} (\frac{\alpha}{q_v \alpha_{rv}})^{-\frac{q_v R_v}{C_v^*}} - C_v^* T_r - q_v R_v T_r + q_v (L_{vr} + L_{fr}) + q_l L_{fr} \\
\label{H-unapprox}
h(p,\eta,q_d,q_v,q_l,q_i) &=& C_p^* T_r \exp (\frac{\eta - \eta_r}{C_p^*}) (\frac{q_d R_d p}{R^* p_{rd}})^{\frac{q_d R_d}{C_p^*}} (\frac{q_v R_v p}{R^* p_{rv}})^{\frac{q_v R_v}{C_p^*}} - C_p^* T_r + q_d R_d T_r + q_v (L_{vr} + L_{fr}) + q_l L_{fr}\\
\label{G-unapprox}
g(p,T,q_d,q_v,q_l,q_i) &=& T (C_p^* - C_p^* \ln \frac{T}{T_r} + q_d R_d \ln \frac{q_d R_d p}{R^* p_{rd}} + q_v R_v \ln \frac{q_v R_v p}{R^* p_{rv}} - \eta_r) - C_p^* T_r + q_d R_d T_r + q_v (L_{vr} + L_{fr}) + q_l L_{fr}\\
\label{F-unapprox}
f(\alpha,T,q_d,q_v,q_l,q_i) &=& T (C_v^* - C_v^* \ln \frac{T}{T_r} + q_d R_d \ln \frac{q_d \alpha_{rd}}{\alpha} + q_v R_v \ln \frac{q_v \alpha_{rv}}{\alpha} - \eta_r) - C_v^* T_r - q_v R_v T_r + q_v (L_{vr} + L_{fr}) + q_l L_{fr}
\end{eqnarray}
where $C_v^* = q_d C_{vd} + q_v C_{vv} + q_l C_{l} + q_i C_i$, $C_p^* = q_d C_{pd} + q_v C_{pv} + q_l C_{l} + q_i C_i$ and $R^* = q_d R_d + q_v R_v$. A detailed derivation for these is found in Appendix \ref{deriv-unapprox}, where the meaning of all symbols (including the latent heat terms $L_{vr}$ and $L_{fr}$) are explained, and the relationship to the potentials from \cite{Thuburn2017,Thuburn2017a,Staniforth2019} is discussed.

From (\ref{U-unapprox}) - (\ref{F-unapprox}) the associated conjugate variables $p$, $\alpha$, $T$ and $\eta$ are given by
\begin{eqnarray}
\label{p-unapprox-U}
p(\alpha,\eta,q_d,q_v,q_l,q_i) &=& -\pp{u}{\alpha} = \frac{R^*}{\alpha} T_r \exp (\frac{\eta - \eta_r}{C_v^*}) (\frac{\alpha}{q_d \alpha_{rd}})^{-\frac{q_d R_d}{C_v^*}} (\frac{\alpha}{q_v \alpha_{rv}})^{-\frac{q_v R_v}{C_v^*}}\\
T(\alpha,\eta,q_d,q_v,q_l,q_i) &=& \pp{u}{\eta} = T_r \exp (\frac{\eta - \eta_r}{C_v^*}) (\frac{\alpha}{q_d \alpha_{rd}})^{-\frac{q_d R_d}{C_v^*}} (\frac{\alpha}{q_v \alpha_{rv}})^{-\frac{q_v R_v}{C_v^*}}\\
\alpha(p,\eta,q_d,q_v,q_l,q_i) &=& \pp{h}{p} = \frac{R^*}{p} T_r \exp (\frac{\eta - \eta_r}{C_p^*}) (\frac{q_d R_d p}{R^* p_{rd}})^{\frac{q_d R_d}{C_p^*}} (\frac{q_v R_v p}{R^* p_{rv}})^{\frac{q_v R_v}{C_p^*}}\\ 
\label{T-H-unapprox}
T(p,\eta,q_d,q_v,q_l,q_i) &=& \pp{h}{\eta} =  T_r \exp (\frac{\eta - \eta_r}{C_p^*}) (\frac{q_d R_d p}{R^* p_{rd}})^{\frac{q_d R_d}{C_p^*}} (\frac{q_v R_v p}{R^* p_{rv}})^{\frac{q_v R_v}{C_p^*}} \\
\alpha(p,T,q_d,q_v,q_l,q_i) &=& \pp{g}{p} = \frac{R^* T}{p}\\
\label{eta-G-unapprox}
    \eta(p,T,q_d,q_v,q_l,q_i) &=& -\pp{g}{T} = C_p^* \ln \frac{T}{T_r} - R^* \ln \frac{p}{R^*} - q_d R_d \ln \frac{q_d R_d}{p_{rd}} - q_v R_v \ln \frac{q_v R_v}{p_{rv}} + \eta_r\\
p(\alpha,T,q_d,q_v,q_l,q_i) &=& -\pp{f}{\alpha}  = \frac{R^* T}{\alpha} \\
    \label{eta-unapprox-F}
\eta(\alpha,T,q_d,q_v,q_l,q_i) &=& -\pp{f}{T}  = C_v^* \ln \frac{T}{T_r} - q_d R_d \ln \frac{q_d \alpha_{rd}}{\alpha} - q_v R_v \ln \frac{q_v \alpha_{rv}}{\alpha} + \eta_r 
\end{eqnarray}
Note multiple expressions for various conjugate variables, such as $p(\alpha,\eta,q_d,q_v,q_l,q_i)$ and $p(\alpha,T,q_d,q_v,q_l,q_i)$, in terms of different variables. From (\ref{p-unapprox-U}) - (\ref{eta-unapprox-F}) the relationship $p \alpha = R^* T$ is immediately apparent, along with expressions for $u$ and $h$ in terms of $T$:
\begin{eqnarray}
u &=& C_v^* (T-T_r) - q_v R_v T_r + q_v (L_{vr} + L_{fr}) + q_l L_{fr} \label{eqn:Uunapprox} \\
h &=& C_p^* (T-T_r) + q_d R_d T_r + q_v (L_{vr} + L_{fr}) + q_l L_{fr}
\end{eqnarray}
The expressions for $\mu_n$ are quite complicated, and are given in Appendix \ref{chemical-potentials-unapprox}.

\subsection{Potential Temperature}
The most common entropic variable encountered in atmospheric dynamics is the \textit{potential temperature} $\theta$, which can be defined in a general way \cite{Zdunkowski2004}, independent of the specific thermodynamic potential used, either explicitly using (\ref{T-H-unapprox}) as
\begin{equation}
    \theta(\eta, q_d,q_v,q_l,q_i) = T(p_r, \eta, q_d,q_v,q_l,q_i)
\end{equation}
or implicitly using (\ref{eta-G-unapprox}) as
\begin{equation}
    \eta(p,T,q_d,q_v,q_l,q_i) = \eta(p_r, \theta, q_d,q_v,q_l,q_i)
\end{equation}
In both expressions, we have simply replaced actual pressure $p$ by some reference pressure $p_r$. In other words, potential temperature is the temperature of an air parcel if it is moved \textit{adiabatically} (at constant entropy and concentrations) from it's actual pressure to some reference pressure. This is simply a specific instance of the general approach to a \textit{potential} quantity discussed in Appendix \ref{potential-virtual-quantities}. The associated conjugate variable (generalized temperature) for $\theta$ is the \textit{Exner pressure} $\Pi$. 

Using (\ref{eta-G-unapprox}) this gives
\begin{equation}
    \eta(\theta,q_d,q_v,q_l,q_i) = C_p^* \ln \frac{\theta}{T_r} - R^* \ln \frac{p_r}{R^*} - q_d R_d \ln \frac{q_d R_d}{p_{rd}} - q_v R_v \ln \frac{q_v R_v}{p_{rv}} + \eta_r \\
\end{equation}
and therefore (using (\ref{T-H-unapprox}))
\begin{equation}
\theta(\eta, q_d,q_v,q_l,q_i) = T_r \exp (\frac{\eta - \eta_r}{C_p^*}) (\frac{p_r}{R^*})^{\frac{R^*}{C_p^*}} (\frac{q_d R_d}{p_{rd}})^{\frac{q_d R_d}{C_p^*}} (\frac{q_v R_v}{p_{rv}})^{\frac{q_v R_v}{C_p^*}}
\end{equation}
where $\kappa^* = \frac{R^*}{C_p^*}$.

The internal energy $u$ (\ref{U-unapprox}) and enthalpy $h$ (\ref{H-unapprox}) in terms of $\theta$ are given by
\begin{eqnarray}
\label{U-unapprox-theta}
    u(\alpha,\theta,q_d,q_v,q_l,q_i) &=& C_v^* (\theta)^{\gamma^*} (\frac{R^*}{\alpha p_r})^{\delta^*}  - C_v^* T_r - q_v R_v T_r + q_v (L_{vr} + L_{fr}) + q_l L_{fr}\\
\label{H-unapprox-theta}
    h(p,\theta,q_d,q_v,q_l,q_i) &=& C_p^* \theta (\frac{p}{p_r})^{\kappa^*} - C_p^* T_r + q_d R_d T_r + q_v (L_{vr} + L_{fr}) + q_l L_{fr}
\end{eqnarray}
where $\gamma^* = \frac{C_p^*}{C_v^*}$ and $\delta^* = \frac{R^*}{C_v^*}$, and we have $\gamma^* - 1 = \delta^*$ and $\kappa^* \gamma^* = \delta^*$. The expressions (\ref{U-unapprox-theta}) - (\ref{H-unapprox-theta}) are significantly simpler than (\ref{U-unapprox}) - (\ref{H-unapprox}), and lead to simpler expressions for the conjugate variables $p$, $\Pi$ and $\alpha$ (one motivation for using an entropic variable), which are
\begin{eqnarray}
\label{p-U-theta-unapprox}
p(\alpha, \theta, q_d,q_v,q_l,q_i) &=& -\pp{u}{\alpha} = p_r  \left( \frac{R^* \theta}{\alpha p_r}\right)^{\gamma^*}\\
\Pi(\alpha,\theta,q_d,q_v,q_l,q_i) &=& \pp{u}{\theta} = C_p^* \left(\frac{R^* \theta}{\alpha p_r}\right)^{\delta^*} \\
\alpha(p, \theta, q_d,q_v,q_l,q_i) &=& \pp{h}{p} = \frac{R^* \theta}{p} \left(\frac{p}{p_r} \right)^{\kappa^*}\\
\label{Pi-H-theta-unapprox}
\Pi(p,\theta,_d,q_v,q_l,q_i) &=& \pp{h}{\theta} = C_p^* \left(\frac{p}{p_r}\right)^{\kappa^*}
\end{eqnarray}
From (\ref{p-U-theta-unapprox}) - (\ref{Pi-H-theta-unapprox}) we have $\Pi \theta = C_p^* T$ and $p \alpha = \kappa^* \Pi \theta$. The Exner pressure $\Pi$ here differs slightly from the form $\Pi = (\frac{p}{p_r})^{\kappa^*}$ often seen in the literature. We prefer the form above since it is the conjugate variable to $\theta$ defined through $\Pi = \pp{U}{\theta}$ from $U(\alpha, \theta, q_n)$. We also have expressions for $u$ and $h$ in terms of $\theta$ and $\Pi$ as
\begin{eqnarray}
    u &=& \frac{C_v^*}{C_p^*} \Pi \theta - C_v^* T_r - q_v R_v T_r + q_v (L_{vr} + L_{fr}) + q_l L_{fr} \\
    h &=& \theta \Pi - C_p^* T_r + q_d R_d T_r + q_v (L_{vr} + L_{fr}) + q_l L_{fr}
\end{eqnarray}
The expressions for generalized chemical potentials $\xi_n$ are quite complicated, and are given in Appendix \ref{chemical-potentials-unapprox}.

\section{Approximated Thermodynamics of Moist Air}
\label{approx-moist-air}

We now consider two widely used approximations to the thermodynamic potentials (\ref{U-unapprox}) - (\ref{F-unapprox}) above. Both involve modifying the heat capacities $C_v^*$ and $C_p^*$ to remove some of the dependence on water species $q_v,q_l,q_i$ and therefore simplify the conjugate variables and other expressions.

%\todo{ADD DISCUSSION ABOUT MAGNITUDE OF APPROXIMATIONS, AND MAYBE SOME NUMERICAL CALCULATIONS?}

\subsection{Constant $\kappa$ system}
\label{constant-kappa}
To obtain the constant $\kappa$ system, make the substitutions:
\begin{equation}
\label{const-kappa-approx}
C_v^* \rightarrow C_{vd} \frac{R^*}{R_d} \quad\quad\quad C_p^* \rightarrow C_{pd} \frac{R^*}{R_d}
\end{equation}
in the thermodynamic potentials (\ref{U-unapprox}) - (\ref{F-unapprox}), but retain $R^*$. This is equivalent to assuming that $\kappa^* = \kappa_d$, hence the name \textit{constant $\kappa$}. In other words, it amounts to the replacement of $C_{vv}$ with $\frac{C_{vd} R_v}{R_d}$ and $C_{pv}$ with $\frac{C_{pd} R_v}{R_d}$, along with the neglect of $C_l$ and $C_i$. It is interesting to note that even with these replacements we still have relationships between heat capacities at constant pressure and constant volume for both water vapor and moist air; which are given by $\frac{C_{vd} R_v}{R_d} + R_v = \frac{C_{pd} R_v}{R_d}$ and $C_{vd} \frac{R^*}{R_d} +R^* = C_{pd} \frac{R^*}{R_d}$.

After some algebra, using (\ref{const-kappa-approx}) in (\ref{U-unapprox}) - (\ref{F-unapprox}) gives
\begin{eqnarray}
\label{U-const-kappa}
u(\alpha,\eta,q_d,q_v,q_l,q_i) &=& C_{vd} \frac{R^*}{R_d} T_r \exp (\frac{\eta - \eta_r}{C_{vd} \frac{R^*}{R_d}}) (\frac{\alpha}{q_d \alpha_{rd}})^{-\frac{q_d R_d}{C_{vd} \frac{R^*}{R_d}}} (\frac{\alpha}{q_v \alpha_{rv}})^{-\frac{q_v R_v}{C_{vd} \frac{R^*}{R_d}}} - C_{vd} \frac{R^*}{R_d} T_r - q_v R_v T_r + q_v (L_{vr} + L_{fr}) + q_l L_{fr} \\
\label{H-const-kappa}
h(p,\eta,q_d,q_v,q_l,q_i) &=& C_{pd} \frac{R^*}{R_d} T_r \exp (\frac{\eta - \eta_r}{C_{pd} \frac{R^*}{R_d}}) (\frac{q_d R_d p}{R^* p_{rd}})^{\frac{q_d R_d}{C_{pd} \frac{R^*}{R_d}}} (\frac{q_v R_v p}{R^* p_{rv}})^{\frac{q_v R_v}{C_{pd} \frac{R^*}{R_d}}} - C_{pd} \frac{R^*}{R_d} T_r + q_d R_d T_r + q_v (L_{vr} + L_{fr}) + q_l L_{fr}\\
g(p,T,q_d,q_v,q_l,q_i) &=& T (- C_{pd} \frac{R^*}{R_d} \ln \frac{T}{T_r} + q_d R_d \ln \frac{q_d R_d p}{R^* p_{rd}} + q_v R_v \ln \frac{q_v R_v p}{R^* p_{rv}} - \eta_r) + C_{pd} \frac{R^*}{R_d} (T-T_r) + q_d R_d T_r + q_v (L_{vr} + L_{fr}) + q_l L_{fr}\\
\label{F-const-kappa}
f(\alpha,T,q_d,q_v,q_l,q_i) &=& T (- C_{vd} \frac{R^*}{R_d} \ln \frac{T}{T_r} + q_d R_d \ln \frac{q_d \alpha_{rd}}{\alpha} + q_v R_v \ln \frac{q_v \alpha_{rv}}{\alpha} - \eta_r) - C_{vd} \frac{R^*}{R_d} (T - T_r) - q_v R_v T_r + q_v (L_{vr} + L_{fr}) + q_l L_{fr}
\end{eqnarray}

The associated conjugate variables $p$, $\alpha$, $T$ and $\eta$ for (\ref{U-const-kappa}) - (\ref{F-const-kappa}) are given by
\begin{eqnarray}
p(\alpha,\eta,q_d,q_v,q_l,q_i) &=& -\pp{u}{\alpha} = \frac{R^*}{\alpha} T_r \exp (\frac{\eta - \eta_r}{C_{vd} \frac{R^*}{R_d}}) (\frac{\alpha}{q_d \alpha_{rd}})^{-\frac{q_d R_d}{C_{vd} \frac{R^*}{R_d}}} (\frac{\alpha}{q_v \alpha_{rv}})^{-\frac{q_v R_v}{C_{vd} \frac{R^*}{R_d}}}\\
T(\alpha,\eta,q_d,q_v,q_l,q_i) &=& \pp{u}{\eta} = T_r \exp (\frac{\eta - \eta_r}{C_{vd} \frac{R^*}{R_d}}) (\frac{\alpha}{q_d \alpha_{rd}})^{-\frac{q_d R_d}{C_{vd} \frac{R^*}{R_d}}} (\frac{\alpha}{q_v \alpha_{rv}})^{-\frac{q_v R_v}{C_{vd} \frac{R^*}{R_d}}}\\
\alpha(p,\eta,q_d,q_v,q_l,q_i) &=& \pp{h}{p} = \frac{R^*}{p} T_r \exp (\frac{\eta - \eta_r}{C_{pd} \frac{R^*}{R_d}}) (\frac{q_d R_d p}{R^* p_{rd}})^{\frac{q_d R_d}{C_{pd} \frac{R^*}{R_d}}} (\frac{q_v R_v p}{R^* p_{rv}})^{\frac{q_v R_v}{C_{pd} \frac{R^*}{R_d}}}\\ 
\label{T-H-constant-kappa}
T(p,\eta,q_d,q_v,q_l,q_i) &=& \pp{h}{\eta} =  T_r \exp (\frac{\eta - \eta_r}{C_{pd} \frac{R^*}{R_d}}) (\frac{q_d R_d p}{R^* p_{rd}})^{\frac{q_d R_d}{C_{pd} \frac{R^*}{R_d}}} (\frac{q_v R_v p}{R^* p_{rv}})^{\frac{q_v R_v}{C_{pd} \frac{R^*}{R_d}}} \\
\alpha(p,T,q_d,q_v,q_l,q_i) &=& \pp{g}{p} = \frac{R^* T}{p}\\
\label{eta-G-constant-kappa}
    \eta(p,T,q_d,q_v,q_l,q_i) &=& -\pp{g}{T} = C_{pd} \frac{R^*}{R_d} \ln \frac{T}{T_r} - R^* \ln \frac{p}{R^*} - q_d R_d \ln \frac{q_d R_d}{p_{rd}} - q_v R_v \ln \frac{q_v R_v}{p_{rv}} + \eta_r\\
    p(\alpha,T,q_d,q_v,q_l,q_i) &=& -\pp{f}{\alpha}  = \frac{R^* T}{\alpha} \\
\eta(\alpha,T,q_d,q_v,q_l,q_i) &=& -\pp{f}{T}  = C_{vd} \frac{R^*}{R_d} \ln \frac{T}{T_r} - q_d R_d \ln \frac{q_d \alpha_{rd}}{\alpha} - q_v R_v \ln \frac{q_v \alpha_{rv}}{\alpha} + \eta_r 
\end{eqnarray}

\begin{remark}
As discussed in Section \ref{review-equilbrium-thermo}, we see that $p \alpha = R^* T$ still holds. This is a good demonstration that these sorts of expressions (often referred to as equations of state in the atmospheric dynamics literature) do not contain a complete description of the thermodynamics, since both the unapproximated and constant $\kappa$ systems give the same expression despite having different thermodynamics.
\end{remark}

Additionally, we have expressions for $u$ and $h$ in terms of $T$ as
\begin{eqnarray}
u &=&  C_{vd} \frac{R^*}{R_d} (T - T_r) - q_v R_v T_r + q_v (L_{vr} + L_{fr}) + q_l L_{fr} \label{eqn:Ukappa} \\
h &=&  C_{pd} \frac{R^*}{R_d} (T - T_r) + q_d R_d T_r + q_v L_{vr} + q_v (L_{vr} + L_{fr}) + q_l L_{fr}
\end{eqnarray}
The expressions for $\mu_n$ are quite complicated, and are given in Appendix \ref{chemical-potentials-const-kappa}.

\subsubsection{Potential Temperature}
As for the unapproximated system, define potential temperature $\theta$ either explicitly using (\ref{T-H-constant-kappa}) or implicitly using (\ref{eta-G-constant-kappa}) by replacing $p$ with $p_r$ to obtain
\begin{equation}
    \eta(\theta,q_d,q_v,q_l,q_i) =C_{pd} \frac{R^*}{R_d} \ln \frac{\theta}{T_r} - R^* \ln \frac{p_r}{R^*} - q_d R_d \ln \frac{q_d R_d}{p_{rd}} - q_v R_v \ln \frac{q_v R_v}{p_{rv}} + \eta_r \\
\end{equation}
\begin{equation}
\theta(\eta, q_d,q_v,q_l,q_i) = T_r \exp (\frac{\eta - \eta_r}{C_{pd} \frac{R^*}{R_d}}) (\frac{p_r}{R^*})^{\frac{R_d}{C_{pd}}} (\frac{q_d R_d}{p_{rd}})^{\frac{q_d R_d}{C_{pd} \frac{R^*}{R_d}}} (\frac{q_v R_v}{p_{rv}})^{\frac{q_v R_v}{C_{pd} \frac{R^*}{R_d}}}  
\end{equation}
Now writing $u$ (\ref{U-const-kappa}) and $h$ (\ref{H-const-kappa}) in terms of $\theta$ we get
\begin{eqnarray}
\label{U-const-kappa-theta}
    u(\alpha,\theta,q_d,q_v,q_l,q_i) &=& C_{vd} \frac{R^*}{R_d} (\theta)^{\gamma_d} (\frac{R^*}{\alpha p_r})^{\delta_d} - C_{vd} \frac{R^*}{R_d} T_r - q_v R_v T_r + q_v (L_{vr} + L_{fr}) + q_l L_{fr}\\
    \label{H-const-kappa-theta}
    h(p,\theta,q_d,q_v,q_l,q_i) &=& C_{pd} \frac{R^*}{R_d}  \theta (\frac{p}{p_r})^{\kappa_d} - C_{pd} \frac{R^*}{R_d} T_r + q_d R_d T_r + q_v (L_{vr} + L_{fr}) + q_l L_{fr}
\end{eqnarray}
The conjugate variables $p$, $\Pi$ and $\alpha$ for (\ref{U-const-kappa-theta}) - (\ref{H-const-kappa-theta}) are
\begin{eqnarray}
\label{p-U-theta-const-kappa}
p(\alpha, \theta, q_d,q_v,q_l,q_i) &=& -\pp{u}{\alpha} = p_r  \left( \frac{R^* \theta}{\alpha p_r}\right)^{\gamma_d}\\
\Pi(\alpha,\theta,q_d,q_v,q_l,q_i) &=& \pp{u}{\theta} = C_{pd} \frac{R^*}{R_d} \left(\frac{R^* \theta}{\alpha p_r}\right)^{\delta_d} \\
\alpha(p, \theta, q_d,q_v,q_l,q_i) &=& \pp{h}{p} = \frac{R^* \theta}{p} \left(\frac{p}{p_r} \right)^{\kappa_d} \\
\label{Pi-H-theta-const-kappa}
\Pi(p,\theta,q_d,q_v,q_l,q_i) &=& \pp{h}{\theta} = C_{pd} \frac{R^*}{R_d} (\frac{p}{p_r})^{\kappa_d}
\end{eqnarray}

Although reduced compared to using $\eta$, there is still non-trivial dependence on $q_v$, $q_l$ and $q_i$ in both the thermodynamic potentials and the conjugate variables, through the $R^*$ terms. From (\ref{p-U-theta-const-kappa}) - (\ref{Pi-H-theta-const-kappa}) we have $p \alpha = \kappa_d \Pi \theta$ and $\Pi \theta = \frac{R^*}{\kappa_d} T$. These two expressions also yield $\theta = T (\frac{p_r}{p})^{\kappa_d}$. Expressions for $u$ and $h$ in terms of $\theta$ and $\Pi$ are
\begin{eqnarray}
u &=& \frac{C_{vd}}{C_{pd}} \Pi \theta - C_{vd} \frac{R^*}{R_d} T_r - q_v R_v T_r + q_v (L_{vr} + L_{fr}) + q_l L_{fr} \\
h &=& \theta \Pi - C_{pd} \frac{R^*}{R_d} T_r + q_d R_d T_r + q_v (L_{vr} + L_{fr}) + q_l L_{fr}
\end{eqnarray}
The expressions for $\xi_n$ are quite complicated, and are given in Appendix \ref{chemical-potentials-const-kappa}.

\subsubsection{Virtual Potential Temperature}
Although using potential temperature simplifies the expressions for the thermodynamic potentials and conjugate variables, there is still non-trivial dependence on the moisture variables $q_v$, $q_l$ and $q_i$. However, it is known from previous work \cite{Taylor2020} that for the constant $\kappa$ approximation there must exist a choice of entropic variable that removes this dependence. This turns out to be the \textit{virtual potential temperature} $\theta_v$ (see Appendix \ref{potential-virtual-quantities} for more discussion of \textit{virtual} quantities), defined through
\begin{eqnarray}
\label{alt-theta-v-expr}
\theta_v(\eta,q_d,q_v,q_l,q_i) &=& T_r \exp (\frac{\eta - \eta_r}{C_{pd} \frac{R^*}{R_d}}) (p_r)^{\frac{R_d}{C_{pd}}} (\frac{q_d R_d}{p_{rd}})^{\frac{q_d R_d}{C_{pd} \frac{R^*}{R_d}}} (\frac{q_v R_v}{p_{rv}})^{\frac{q_v R_v}{C_{pd} \frac{R^*}{R_d}}} \frac{1}{R_d}(R^*)^{\frac{C_{vd}}{C_{pd}}}\\
\eta(\theta_v,q_d,q_v,q_l,q_i) &=& C_{pd} \frac{R^*}{R_d} \ln \frac{\theta_v}{T_r} - R^* \ln p_r - q_d R_d \ln \frac{q_d R_d}{p_{rd}} - q_v R_v \ln \frac{q_v R_v}{p_{rv}} + \eta_r + \frac{R^*}{R_d} \left( C_{pd} \ln R_d - C_{vd} \ln R^*\right)
\end{eqnarray}
This is simply the expression for $\theta_v$ in terms of $p$, $\eta$ and $q_n$, which is independent of $p$ for the constant $\kappa$ system (i.e. (\ref{theta-v-expr})), and therefore an entropic variable. The associated conjugate variable turns out to be, as might be expected, the \textit{virtual Exner pressure} $\Pi_v$. 
%in that the $R^* \ln R^*$ term in $\eta(\theta, q_d,q_v,q_l,q_i)$ is replaced with $\frac{R^*}{R_d} \left( C_{pd} \ln R_d - C_{vd} \ln R^*\right)$ (and a similar replacement in $\theta_v$). 

Now writing $u$ (\ref{U-const-kappa}) and $h$ (\ref{H-const-kappa}) in terms of $\theta_v$ we get
\begin{eqnarray}
\label{U-const-kappa-thetav}
    u(\alpha,\theta_v,q_d,q_v,q_l,q_i) &=& C_{vd} (\theta_v)^{\gamma_d} (\frac{R_d}{\alpha p_r})^{\delta_d} - C_{vd} \frac{R^*}{R_d} T_r - q_v R_v T_r + q_v (L_{vr} + L_{fr}) + q_l L_{fr}\\
    \label{H-const-kappa-thetav}
    h(p,\theta_v,q_d,q_v,q_l,q_i) &=& C_{pd} \theta_v (\frac{p}{p_r})^{\kappa_d} - C_{pd} \frac{R^*}{R_d} T_r + q_d R_d T_r + q_v (L_{vr} + L_{fr}) + q_l L_{fr}
\end{eqnarray}
Crucially, almost all of the dependence of $U$ and $H$ on $q_d,q_v,q_l,q_i$ has been absorbed into $\theta_v$, other than trivial linear dependence that will affect only $\xi_n$.

The conjugate variables $p$, $\Pi_v$ and $\alpha$ for (\ref{U-const-kappa-thetav}) - (\ref{H-const-kappa-thetav}) are
\begin{eqnarray}
\label{p-U-thetav-const-kappa}
p(\alpha, \theta_v, q_d,q_v,q_l,q_i) &=& -\pp{u}{\alpha} = p_r  \left( \frac{R_d \theta_v}{\alpha p_r}\right)^{\gamma_d}\\
\Pi_v(\alpha,\theta_v,q_d,q_v,q_l,q_i) &=& \pp{u}{\theta_v} = C_{pd} \left(\frac{R_d \theta_v}{\alpha p_r}\right)^{\delta_d} \\
\alpha(p, \theta_v, q_d,q_v,q_l,q_i) &=& \pp{h}{p} = \frac{R_d \theta_v}{p} \left(\frac{p}{p_r} \right)^{\kappa_d} \\
\label{Piv-H-thetav-const-kappa}
\Pi_v(p,\theta_v,_d,q_v,q_l,q_i) &=& \pp{h}{\theta_v} = C_{pd} (\frac{p}{p_r})^{\kappa_d}
\end{eqnarray}
Due to the absorption of most of the dependence on $q_d,q_v,q_l,q_i$ into $\theta_v$, the expressions for $p$, $\Pi_v$ and $\alpha$ are independent of $q_d,q_v,q_l,q_i$. This is extremely useful, because with an appropriate choice of predicted variables this means the adiabatic (i.e. reversible) dynamics of the water species will decouple from the rest of the dynamics. This permits utilizing different spatial and temporal numerics for the two sets of variables. An example of this for the HOMME-NH dynamical core is found in \cite{Taylor2020}.

From (\ref{p-U-thetav-const-kappa}) - (\ref{Piv-H-thetav-const-kappa}) we have $p \alpha = \kappa_d \Pi_v \theta_v$ and $\Pi_v \theta_v = \frac{R^*}{\kappa_d} T$. The former is usually taken as the starting point for the definition of $\theta_v$.  These two expressions also yield  $\theta_v = \frac{R^*}{R_d} T (\frac{p_r}{p})^{\kappa_d} = T_v (\frac{p_r}{p})^{\kappa_d}$ for virtual temperature $T_v = T \frac{R^\star}{R_d}$, which is slightly different than the naively expected $\theta_v = T (\frac{p_r}{p})^{\kappa_d}$. Expressions for $u$ and $h$ in terms of $\theta_v$ and $\Pi_v$ are
\begin{eqnarray}
u &=& \frac{C_{vd}}{C_{pd}} \Pi_v \theta_v - C_{vd} \frac{R^*}{R_d} T_r - q_v R_v T_r + q_v (L_{vr} + L_{fr}) + q_l L_{fr} \\
h &=& \theta_v \Pi_v - C_{pd} \frac{R^*}{R_d} T_r + q_d R_d T_r + q_v (L_{vr} + L_{fr}) + q_l L_{fr}
\end{eqnarray}
The expressions for $\xi_n$ are quite complicated, and are given in Appendix \ref{chemical-potentials-const-kappa}.

\subsection{Dry heat capacities system}
\label{dry-heat-capacities}

Another approximation often used is to assume that all of the heat capacities of the various species are the same (see Appendix \ref{common-thermo-quantities} for a general definition of heat capacity), such as in \cite{boville03}. Usually these are taken to be the heat capacity of dry air, although this is not required. We will retain the notation $C_{vd}$ and $C_{pd}$. To accomplish this, in the thermodynamic potentials (\ref{U-unapprox}) - (\ref{F-unapprox}) make the substitution:
\begin{equation}
C_v^* \rightarrow C_{vd} \quad\quad\quad C_p^* \rightarrow C_{pd}
\end{equation}
but retain $R^*$. In this case, we no longer have the relationship $C_v + R^* = C_p$, which leads to interesting artifacts when using potential temperature as an entropic variable.

This gives (after some algebra)
\begin{eqnarray}
\label{U-dry-c}
u(\alpha,\eta,q_d,q_v,q_l,q_i) &=& C_{vd} T_r \exp (\frac{\eta - \eta_r}{C_{vd}}) (\frac{\alpha}{q_d \alpha_{rd}})^{-\frac{q_d R_d}{C_{vd}}} (\frac{\alpha}{q_v \alpha_{rv}})^{-\frac{q_v R_v}{C_{vd}}} - C_{vd} T_r - q_v R_v T_r + q_v (L_{vr} + L_{fr}) + q_l L_{fr} \\
    \label{H-dry-c}
h(p,\eta,q_d,q_v,q_l,q_i) &=& C_{pd} T_r \exp (\frac{\eta - \eta_r}{C_{pd}}) (\frac{q_d R_d p}{R^* p_{rd}})^{\frac{q_d R_d}{C_{pd}}} (\frac{q_v R_v p}{R^* p_{rv}})^{\frac{q_v R_v}{C_{pd}}} - C_{pd} T_r + q_d R_d T_r + q_v (L_{vr} + L_{fr}) + q_l L_{fr}\\
g(p,T,q_d,q_v,q_l,q_i) &=& T (C_{pd} - C_{pd} \ln \frac{T}{T_r} + q_d R_d \ln \frac{q_d R_d p}{R^* p_{rd}} + q_v R_v \ln \frac{q_v R_v p}{R^* p_{rv}} - \eta_r) - C_{pd} T_r + q_d R_d T_r + q_v (L_{vr} + L_{fr}) + q_l L_{fr}\\
\label{F-dry-c}
f(\alpha,T,q_d,q_v,q_l,q_i) &=& T (C_{vd} - C_{vd} \ln \frac{T}{T_r} + q_d R_d \ln \frac{q_d \alpha_{rd}}{\alpha} + q_v R_v \ln \frac{q_v \alpha_{rv}}{\alpha} - \eta_r) - C_{vd} T_r - q_v R_v T_r + q_v (L_{vr} + L_{fr}) + q_l L_{fr}
\end{eqnarray}

The associated conjugate variables $p$, $\alpha$, $T$ and $\eta$ for (\ref{U-dry-c}) - (\ref{F-dry-c}) are given by
\begin{eqnarray}
\label{p-U-dry-c}
p(\alpha,\eta,q_d,q_v,q_l,q_i) &=& -\pp{u}{\alpha} = \frac{R^*}{\alpha} T_r \exp (\frac{\eta - \eta_r}{C_{vd}}) (\frac{\alpha}{q_d \alpha_{rd}})^{-\frac{q_d R_d}{C_{vd}}} (\frac{\alpha}{q_v \alpha_{rv}})^{-\frac{q_v R_v}{C_{vd}}}\\
T(\alpha,\eta,q_d,q_v,q_l,q_i) &=& \pp{u}{\eta} = T_r \exp (\frac{\eta - \eta_r}{C_{vd}}) (\frac{\alpha}{q_d \alpha_{rd}})^{-\frac{q_d R_d}{C_{vd}}} (\frac{\alpha}{q_v \alpha_{rv}})^{-\frac{q_v R_v}{C_{vd}}}\\
\alpha(p,\eta,q_d,q_v,q_l,q_i) &=& \pp{h}{p} = \frac{R^*}{p} T_r \exp (\frac{\eta - \eta_r}{C_{pd}}) (\frac{q_d R_d p}{R^* p_{rd}})^{\frac{q_d R_d}{C_{pd}}} (\frac{q_v R_v p}{R^* p_{rv}})^{\frac{q_v R_v}{C_{pd}}}\\ 
\label{T-H-dry-c}
T(p,\eta,q_d,q_v,q_l,q_i) &=& \pp{h}{\eta} =  T_r \exp (\frac{\eta - \eta_r}{C_{pd}}) (\frac{q_d R_d p}{R^* p_{rd}})^{\frac{q_d R_d}{C_{pd}}} (\frac{q_v R_v p}{R^* p_{rv}})^{\frac{q_v R_v}{C_{pd}}} \\
\alpha(p,T,q_d,q_v,q_l,q_i) &=& \pp{g}{p} = \frac{R^* T}{p}\\
\label{eta-G-dry-c}
    \eta(p,T,q_d,q_v,q_l,q_i) &=& -\pp{g}{T} = C_{pd} \ln \frac{T}{T_r} - R^* \ln \frac{p}{R^*} - q_d R_d \ln \frac{q_d R_d}{p_{rd}} - q_v R_v \ln \frac{q_v R_v}{p_{rv}} + \eta_r\\
p(\alpha,T,q_d,q_v,q_l,q_i) &=& -\pp{f}{\alpha}  = \frac{R^* T}{\alpha} \\
\label{eta-F-dry-c}
\eta(\alpha,T,q_d,q_v,q_l,q_i) &=& -\pp{f}{T}  = C_{vd} \ln \frac{T}{T_r} - q_d R_d \ln \frac{q_d \alpha_{rd}}{\alpha} - q_v R_v \ln \frac{q_v \alpha_{rv}}{\alpha} + \eta_r 
\end{eqnarray}
Again, from (\ref{p-U-dry-c}) - (\ref{eta-F-dry-c}) the relationship $p \alpha = R^* T$ still holds, along with expressions for $u$ and $h$ in terms of $T$
\begin{eqnarray}
u &=& C_{vd} (T-T_r) - q_v R_v T_r + q_v (L_{vr} + L_{fr}) + q_l L_{fr} \label{eqn:Udry} \\ 
h &=& C_{pd} (T-T_r) + q_d R_d T_r + q_v (L_{vr} + L_{fr}) + q_l L_{fr}
\end{eqnarray}
The expressions for $\mu_n$ are quite complicated, and are given in Appendix \ref{chemical-potentials-dry-heat}.

\subsubsection{Potential Temperature}
As done for the previous two systems, define potential temperature $\theta$ either explicitly using (\ref{T-H-dry-c}) or implicitly using (\ref{eta-G-dry-c}) by replacing $p$ with $p_r$ to obtain
\begin{equation}
    \eta(\theta,q_d,q_v,q_l,q_i) =C_{pd} \ln \frac{\theta}{T_r} - R^* \ln \frac{p_r}{R^*} - q_d R_d \ln \frac{q_d R_d}{p_{rd}} - q_v R_v \ln \frac{q_v R_v}{p_{rv}} + \eta_r \\
\end{equation}
\begin{equation}
\theta(\eta, q_d,q_v,q_l,q_i) = T_r \exp (\frac{\eta - \eta_r}{C_{pd}}) (\frac{p_r}{R^*})^{\frac{R_d}{C_{pd}}} (\frac{q_d R_d}{p_{rd}})^{\frac{q_d R_d}{C_{pd}}} (\frac{q_v R_v}{p_{rv}})^{\frac{q_v R_v}{C_{pd}}}  
\end{equation}
Now writing $u$ (\ref{U-dry-c}) and $h$ (\ref{H-dry-c}) in terms of $\theta$ we get
\begin{eqnarray}
\label{U-dry-c-theta}
    u(\alpha,\theta,q_d,q_v,q_l,q_i) &=& C_{vd} (\theta)^{\gamma_d} (\frac{R^*}{\alpha p_r})^{\frac{R^*}{C_{vd}}} (T_r)^{\frac{R^* - R_d}{C_{vd}}} - C_{vd} T_r - q_v R_v T_r + q_v (L_{vr} + L_{fr}) + q_l L_{fr}\\
    \label{H-dry-c-theta}
    h(p,\theta,q_d,q_v,q_l,q_i) &=& C_{pd}  \theta (\frac{p}{p_r})^{\frac{R^*}{C_{vd}}} - C_{pd} T_r + q_d R_d T_r + q_v (L_{vr} + L_{fr}) + q_l L_{fr}
\end{eqnarray}
The strange term $(T_r)^{\frac{R^* - R_d}{C_{vd}}}$ that appears in the $u$ equation is a consequence of the fact $C_v + R^* \neq C_p$.

The conjugate variables $p$, $\Pi$ and $\alpha$ for (\ref{U-dry-c-theta}) - (\ref{H-dry-c-theta}) are
\begin{eqnarray}
\label{p-U-theta-dry-c}
p(\alpha, \theta, q_d,q_v,q_l,q_i) &=& -\pp{u}{\alpha} = \frac{R^*}{\alpha} (\theta)^{\gamma_d} \left( \frac{R^* \theta}{\alpha p_r}\right)^{\frac{R^*}{C_{vd}}} (T_r)^{\frac{R^* - R_d}{C_{vd}}}\\
\Pi(\alpha,\theta,q_d,q_v,q_l,q_i) &=& \pp{u}{\theta} = C_{pd} (\theta)^{\delta_d} \left(\frac{R^*}{\alpha p_r}\right)^{\frac{R^*}{C_{vd}}} (T_r)^{\frac{R^* - R_d}{C_{vd}}} \\
\alpha(p, \theta, q_d,q_v,q_l,q_i) &=& \pp{h}{p} = \frac{R^* \theta}{p} \left(\frac{p}{p_r} \right)^{\frac{R^*}{C_{pd}}} \\
\label{Pi-H-theta-dry-c}
\Pi(p,\theta,q_d,q_v,q_l,q_i) &=& \pp{h}{\theta} = C_{pd} (\frac{p}{p_r})^{\frac{R^*}{C_{pd}}}
\end{eqnarray}
There is some simplification and reduction of dependence on moisture variables, but $u$, $h$ and the associated conjugate variables still have non-trivial dependence through $R^*$. From (\ref{p-U-theta-dry-c}) - (\ref{Pi-H-theta-dry-c}) we have $p \alpha = \frac{R^*}{C_{pd}} \Pi \theta$ and $\Pi \theta = C_{pd} T$. These two expressions also yield $\theta = T (\frac{p_r}{p})^{\frac{R^*}{C_{pd}}}$. Expressions for $u$ and $h$ in terms of $\theta$ and $\Pi$ are
\begin{eqnarray}
u &=& \frac{C_{vd}}{C_{pd}} \Pi \theta - C_{vd} T_r - q_v R_v T_r + q_v (L_{vr} + L_{fr}) + q_l L_{fr} \\
h &=& \theta \Pi - C_{pd} T_r + q_d R_d T_r + q_v (L_{vr} + L_{fr}) + q_l L_{fr}
\end{eqnarray}
The expressions for $\xi_n$ are quite complicated, and are given in Appendix \ref{chemical-potentials-dry-heat}.

\subsection{Quantifying Errors}
\label{inconsistency}
Detailed discussion about both inconsistency and approximation errors in thermodynamics for coupled climate models can be found in \cite{Lauritzen2020}, in this section we provide only a few small examples. Specifically, we estimate discrepancies between the dry heat capacities, constant $\kappa$ and the unapproximated systems using a typical value for water vapor concentration in the tropics, $q_{v}=0.01$ \cite{Vallis}. For thermodynamic constants, we take approximate values $C_{pd} = 1006$, $C_{pv} = 1872$, $C_{vd}=719$, $C_{vv}=1410$, $R_d=287$, and $R_{v}=462$, all in units \si{J~K^{-1}~ kg^{-1}} \cite{E1994}. In what follows, for simplicity we will also assume that there is only one water form in the atmosphere (water vapor), since it is the most dominant form; and therefore $q_l = q_i = 0$. 

Let us start by considering the change in internal energy $\Delta u$ due to a change $\Delta T$ in temperature:
\begin{equation}
\Delta u_{unapprox} = C_v^* \Delta T\quad\quad
\Delta u_{constant-\kappa} = C_{vd} \frac{R^*}{R_d} \Delta T\quad\quad
\Delta u_{dry} = C_{vd} \Delta T
\end{equation}
With $q_v = 0.01$, we have
\begin{eqnarray}
C_v^* \approx 725.91 \si{J~K^{-1}~ kg^{-1}} \quad\quad
C_{vd} \frac{R^*}{R_d} \approx 723.84 \si{J~K^{-1}~ kg^{-1}} \quad\quad
C_{vd} \approx 719 \si{J~K^{-1}~ kg^{-1}}
\end{eqnarray}
So the constant $\kappa$ system will underestimate the change in $u$ by 0.35\%, while the dry heat capacities system will underestimate the change by 0.95\%. This is an almost three times larger error, although still numerically quite small.

However, such computations may not be too informative in applications like global climate system modelling. Instead, consider global climatological means of water enthalpy fluxes for precipitation and evaporation at the atmosphere--ocean interface for the Energy Exascale Earth System Model (E3SM) \cite{E3SM-coupled} or the Community Earth System Model (CESM) \cite{cesm2}, which both currently use the dry heat capacities approach. The enthalpy fluxes for evaporation and precipitation are computed using specific heat capacity of the dry air, $C_{pd}\simeq1006$ \si{J~K^{-1}~ kg^{-1}}, equal approximately 10 \si{W~m^{-2}} each as a global mean average, and are largely based on fluctuations of water vapor. In the unapproximated system, the same fluxes would be computed with $C_{pv}\simeq1872$ \si{J~K^{-1}~ kg^{-1}}.  Therefore, the dry heat capacities system underestimates water energy fluxes in the global model by almost a factor of 2. For global climate simulations, it is a significant difference. On the other hand, in the constant $\kappa$ system the enthalpy fluxes would be computed with the specific heat capacity of water vapor corresponding to this system, $C_{pd}\frac{R_v}{R_d}\simeq 1618$ \si{J~K^{-1}~ kg^{-1}}, and enthalpy fluxes in this system would differ from the fluxes in the unapproximated system only by 16\%. 

\section{Conclusions}
\label{conclusions}

This paper has presented complete sets of thermodynamic potentials in terms of their natural variables for three systems describing moist air: \textit{unapproximated}, constant $\kappa$ and dry heat capacities; along with the associated thermodynamic quantities and relationships derived consistently from them. It is intended as a comprehensive reference for the use of thermodynamic potentials in atmospheric modelling, especially for the three systems considered here. An interesting future direction would be going the other direction and relaxing some of the approximations made here. For example, introducing temperature-dependent heat capacities, condensates with volume, mixed (non-pure) condensates (including treatment of hydrometeor populations) and some type of equilibrium between water phases. This would be useful for more sophisticated atmospheric models, such as high-resolution large eddy simulations or models for exoplanetary atmospheres; and in advanced physics parameterizations.

\section{Acknowledgements}
We thank two anonymous reviewers for their comments and suggestions, which substantially improved the presentation and content of this article.

Sandia National Laboratories is a multimission laboratory managed and operated by National Technology \& Engineering Solutions of Sandia, LLC, a wholly owned subsidiary of Honeywell International Inc., for the U.S. Department of Energy's National Nuclear Security Administration under contract DE-NA0003525. This paper describes objective technical results and analysis. Any subjective views or opinions that might be expressed in the paper do not necessarily represent the views of the U.S. Department of Energy or the United States Government.

This research was supported by the Exascale Computing Project (17-SC-20-SC), a collaborative effort of two U.S. Department of Energy organizations (Office of Science and the National Nuclear Security Administration) responsible for the planning and preparation of a capable exascale ecosystem, including software, applications, hardware, advanced system engineering, and early testbed platforms, in support of the nation's exascale computing imperative.

This research was supported as part of the Energy Exascale Earth System Model (E3SM) project, funded by the U.S. Department of Energy, Office of Science, Office of Biological and Environmental Research.

This work was supported by the U.S. Department of Energy, Office of Science, Advanced Scientific Computing Research (ASCR) Program and Biological and Environmental Research (BER) Program under a Scientific Discovery through Advanced Computing (SciDAC 4) BER partnership pilot project.

\bibliographystyle{abbrv}
\bibliography{main}

\appendix

\section{Deriving \textit{Unapproximated} Thermodynamic Potentials}
\label{deriv-unapprox}

Recall we are considering a mixture of dry air and the three phases of water: water vapor, (cloud) liquid water and (cloud) ice; with concentrations denoted by $q_d$, $q_v$, $q_l$ and $q_i$. The gaseous components $q_d$ and $q_v$ are assumed to be perfect ideal gases with temperature-independent heat capacities at constant volume $C_{vd}$ and $C_{vv}$, heat capacities at constant pressure $C_{pd}$ and $C_{pv}$; and gas constants $R_d$ and $R_v$. The condensed components $q_l$ and $q_i$ are assumed to be incompressible with temperature-independent heat capacities $C_l$ and $C_i$ and to appear as pure substances, and we neglect the volume occupied by the condensates. We additionally assume that the different water phases behave as separate thermodynamic substances, and therefore is no equilibrium between phases, so we can capture super saturation and other out of equilibrium situations. This means that Gibbs phase rule does not have to satisfied.

The assumption of zero volume, incompressible condensates along with ideal gas behaviour for dry air and water vapor means that the partial pressures for the gaseous components are determined by Dalton's law:
\begin{equation}
\label{daltons}
    p \alpha = (p_d + p_v) \alpha = q_d R_d T + q_v R_v T = R^* T
\end{equation}
with $R^* = q_d R_d + q_v R_v$, where $p_d = \frac{q_d R_d}{R^*} p$ and $p_v = \frac{q_v R_v}{R^*} p$ are the partial pressures of dry air and water vapor.

Under the assumption of a single temperature $T$ for all components along with zero volume, incompressible condensates, the total Gibbs function can be written as the concentration-weighted sum of Gibbs function for each component \cite{Zdunkowski2004} plus some additional terms related to the latent heat of the vapor and liquid phases of water, using the common $T$ and the relevant partial pressures from (\ref{daltons}). 

%Since the partial pressures depend only on $q_d$ and $q_v$, this means that the Gibbs function for each component $k$ is independent of the other $q_n's$. This is not true for other thermodynamic potentials, and significantly simplifies the resulting derivation.

Specifically, we start with the well-known Gibbs function for a single component perfect ideal gas with heat capacities $C_v$ and $C_p$, and gas constant $R$ \cite{Zdunkowski2004}:
\begin{equation}
\label{ideal-gas-gibbs}
    g_{g}(p, T) = C_p (T - T_r) + T( - C_p \ln \frac{T}{T_r} + R \ln \frac{p}{p_r} - \eta_0)
\end{equation}
where $\eta_0$ is the specific entropy at $T_r$ and $p_r$ (with $p_r \alpha_r = R T_r$). The pressure here will become the partial pressure for the component in the full expression. Note that this differs from the Gibbs function for an ideal gas found in \cite{Thuburn2017} by a function of the form $A + B T$. This is acceptable, since this changes only the zero of entropy $\eta = -\pp{g}{T}$. Here we have normalized $g$ such that $g = - T_r \eta_0$ at $T =T_r$ and $p = p_r$, and therefore $\eta = \eta_0$ at $T=T_r$ and $p=p_r$. 

For a pure, incompressible, zero-volume condensate with heat capacity $C$ the Gibbs function is \cite{Zdunkowski2004}:
\begin{equation}
\label{pure-gibbs}
    g_{c}(p, T) = C (T - T_r) + T( - C \ln \frac{T}{T_r} - \eta_0)
\end{equation}
where $\eta_0$ is the specific entropy at $T_r$. There is no dependence on $p_r$ since we have assumed incompressibility, and $g$ is normalized such that $g =- T_r \eta_0$ at $T =T_r$ and again therefore $\eta = \eta_0$ at $T=T_r$ and $p=p_r$. As before this differs from the Gibbs function for a condensate found in \cite{Thuburn2017} by a function of the form $A + B T$.

Using (\ref{ideal-gas-gibbs}) and (\ref{pure-gibbs}), we can write the partial Gibbs functions for each component as
\begin{eqnarray}
\label{gibbs-dry}
g_d(p,T,q_d) &=& C_{pd} (T - T_r) + T( - C_{pd} \ln \frac{T}{T_r} + R_d \ln \frac{p_d}{p_{rd}} - \eta_{rd}) + R_d T_r\\
g_v(p,T,q_v) &=& C_{pv} (T - T_r) + T( - C_{pv} \ln \frac{T}{T_r} + R_v \ln \frac{p_v}{p_{rv}} - \eta_{rv}) + L_{fr} + L_{vr}\\
g_l(p,T,q_l) &=& C_l (T - T_r) + T( - C_l \ln \frac{T}{T_r} - \eta_{rl}) + L_{fr} \\
\label{gibbs-ice}
g_i(p,T,q_i) &=& C_i (T - T_r) + T( - C_i \ln \frac{T}{T_r} - \eta_{ri})
\end{eqnarray}
where $p_{rd} \alpha_{rd} = R_d T_r$, $p_{rv} \alpha_{rv} = R_v T_r$, $L_{fr}$ is the latent heat of freezing at $T_r$ and $L_{vr}$ is the latent heat of vaporization at $T_r$; recalling $p_d = \frac{q_d R_d}{R^*} p$ and $p_v = \frac{q_v R_v}{R^*} p$. Note that some care must be taken in the choice of $p_{rd}$ and $p_{rv}$ in order to obtain the correct saturation vapor pressure. The latent heat terms are required in order to obtain correct expressions for the equilibrium between phases of water, although this is not treated here. They are also dependent on the choice of reference state, as discussed in \cite{Lauritzen2020}. The last term ($R_d T_r$) in the $g_d$ equation simply shifts the zero of the partial Gibbs function for dry air, and amounts only to a change in the zero of chemical potential $\mu_d$. Now the total Gibbs function is just the concentration weighted sum of the partial Gibbs functions (\ref{gibbs-dry}) - (\ref{gibbs-ice}):
\begin{equation}
g = q_d g_d + q_v g_v + q_l g_l + q_i g_i 
\end{equation}
% In general, U and F both get terms of the form -R^* Tr + G-TERM, while G and H have terms of the form G-TERM. In here we have chosen G-TERM = q_d R_d T_r, but we could make other choices as well.

%As discussed above, these partial Gibbs's functions depend at most on the concentrations of that component. 

Putting this all together gives
\begin{equation}
\label{full-gibbs}
    g(p,T,q_d,q_v,q_l,q_i) = T (C_p^* - C_p^* \ln \frac{T}{T_r} + q_d R_d \ln \frac{q_d R_d p}{R^* p_{rd}} + q_v R_v \ln \frac{q_v R_v p}{R^* p_{rv}} - \eta_r) - C_p^* T_r + q_v (L_{fr} + L_{vr}) + q_l L_{fr} + q_d R_d T_r\\
\end{equation}
where $C_p^* = q_d C_{pd} + q_v C_{pv} + q_l C_l + q_i C_i$, $C_v^* = q_d C_{vd} + q_v C_{vv} + q_l C_l + q_i C_i$ and $\eta_r = q_d \eta_{rd} + q_v \eta_{rv} + q_l \eta_{rl} + q_i \eta_{ri}$, with $C_p^* = C_v^* + R^*$. This is (\ref{G-unapprox}). As in the partial Gibbs function cases, this (assuming $q_i = 0$) differs from the moist air Gibbs function found in \cite{Thuburn2017} by a function of the form $A + B T + \sum_n C_n q_n$, where the last two terms just shift the zeros of entropy $\eta$ and chemical potentials $\mu_n$, respectively.

From (\ref{full-gibbs}) along with the expressions of the conjugate variables $\alpha(p, T, q_n)$ and $\eta(p,T,q_n)$, it is possible to derive all the other thermodynamic potentials. For example, $\eta(p,T,q_n)$ can be solved to yield $T(p, \eta, q_n)$, and then $h(p,\eta,q_n) = g(p, T(p, \eta, q_n), q_n) + \eta T(p, \eta, q_n)$ i.e. (\ref{H-unapprox}). Similar manipulations give $u(\alpha, \eta, q_n)$ (\ref{U-unapprox}) and $f(\alpha, T, q_n)$ (\ref{F-unapprox}), and are left as an exercise for the interested reader.

\section{Potential and Virtual Quantities}
\label{potential-virtual-quantities}

\subsection{Potential Quantities}
A \textit{potential} quantity (denoted here with $X_p$) is the value a thermodynamic quantity would have if pressure $p$ changed to some reference pressure $p_r$ holding specific entropy $\eta$ and concentrations $q_n$ fixed (an adiabatic process). In other words, they are defined through a relationship of the form
\begin{equation}
    X_p(\eta,q_n) = X(p_r,\eta,q_n)
\end{equation}
where $X(p,\eta,q_n)$ is the original variable.  Note that this definition implies that potential variables are also entropic variables. For example, potential temperature $\theta$ is defined through
\begin{equation}
    \theta(\eta,q_n) = T(p_r,\eta,q_n) 
\end{equation}
In other words, it is the temperature $T$ if $p \rightarrow p_r$ while holding $\eta$ and $q_n$ fixed.

\subsection{Virtual Quantities}
A \textit{virtual} quantity (denoted here with $X_v$) is the value a thermodynamic quantity would have if the concentration of a given component $q_d$ changed to $1$ while all other concentrations $q_s$ went to zero, holding pressure $p$ and specific volume $\alpha$ fixed. In other words, they are defined through a relationship of the form
\begin{equation}
    X_v(p, \alpha) = X(p, \alpha, q_d=1, q_s=0)
\end{equation}
where $X(p, \alpha, q_d, q_s)$ is the original quantity. Note that these are not entropic variables, in general. For atmospheric dynamics, the usual case has $q_d$ as dry air and $q_s$ as water substances, and so a virtual quantity represents the value of a quantity a dry parcel with the same pressure and specific volume as a moist sample would have. For example, virtual temperature is defined (implicitly) through
\begin{equation}
    p(\alpha, T, q_d, q_s) = p(\alpha, T_v, 1, 0)
\end{equation}

\subsubsection{Examples}
To make this more concrete, let us consider \textit{virtual temperature }$T_v$, \textit{virtual potential temperature} $\theta_v$ and \textit{virtual Exner pressure} $\Pi_v$.

\paragraph{Unapproximated system}
Start by considering the \textit{unapproximated} system introduced in Section \ref{unapprox-moist-air}. In this case we have
\begin{eqnarray}
         p(\alpha, T, q_d, q_v, q_l, q_i) &=& \frac{R^* T}{\alpha} \\
         p(\alpha, \theta, q_d, q_v, q_l, q_i) &=& p_r  \left( \frac{R^* \theta}{\alpha p_r}\right)^{\gamma^*} \\
       p(\alpha, \Pi, q_d, q_v, q_l, q_i) &=& p_r (\frac{\Pi}{C_p^*})^{\frac{C_p^*}{R^*}}
\end{eqnarray}
and therefore
\begin{eqnarray}
    T_v(\alpha, p) &=& \frac{p \alpha}{R_d} \\
    \theta_v(\alpha, p) &=&  (\frac{p_r}{p})^{\kappa_d} \frac{p \alpha}{R_d}\\
\Pi_v(\alpha, p) &=& C_{pd} (\frac{p}{p_r})^{\kappa_d}
\end{eqnarray}
An interesting question is whether $\theta_v$ is an entropic variable. This can be investigated by using the expression for $\alpha = \alpha(p,\eta,q_n)$ to obtain $\theta_v(p,\eta,q_n)$. Doing so gives
\begin{equation}
    \theta_v(p,\eta,q_n) = (\frac{p_r}{p})^{\kappa_d} (\frac{p}{R^*})^{\kappa^*} \frac{R^*}{R_d} T_r \exp^{\frac{\eta - \eta_r}{C_p^*}} \left( \frac{q_d R_d}{p_{rd}}\right)^{\frac{q_d R_d}{C_p^*}}  \left( \frac{q_v R_v}{p_{rv}}\right)^{\frac{q_v R_v}{C_p^*}} 
\end{equation}
and therefore we see that $\theta_v$ is not an entropic variable for the unapproximated system.

\paragraph{Constant $\kappa$ system}
Next consider the constant $\kappa$ system introduced in Section \ref{constant-kappa}. In this case we have
\begin{eqnarray}
         p(\alpha, T, q_d, q_v, q_l, q_i) &=& \frac{R^* T}{\alpha} \\
         p(\alpha, \theta, q_d, q_v, q_l, q_i) &=& p_r  \left( \frac{R^* \theta}{\alpha p_r}\right)^{\gamma_d} \\
       p(\alpha, \Pi, q_d, q_v, q_l, q_i) &=& p_r (\frac{\Pi}{C_{pd} \frac{R^*}{R_d}})^{\frac{C_{pd}}{R_d}}
\end{eqnarray}
and therefore
\begin{eqnarray}
    T_v(\alpha, p) &=& \frac{p \alpha}{R_d} \\
    \theta_v(\alpha, p) &=&  (\frac{p_r}{p})^{\kappa_d} \frac{p \alpha}{R_d}\\
\Pi_v(\alpha, p) &=& C_{pd} (\frac{p}{p_r})^{\kappa_d}
\end{eqnarray}
Using the expression for $\alpha = \alpha(p,\eta,q_n)$ to obtain $\theta_v(p,\eta,q_n)$ gives
\begin{equation}
\label{theta-v-expr}
    \theta_v(p,\eta,q_n) = (\frac{p_r}{R^*})^{\kappa_d} \frac{R^*}{R_d} T_r \exp^{\frac{\eta - \eta_r}{C_{pd} \frac{R^*}{R_d}}} \left( \frac{q_d R_d}{p_{rd}}\right)^{\frac{q_d R_d}{C_{pd} \frac{R^*}{R_d}}}  \left( \frac{q_v R_v}{p_{rv}}\right)^{\frac{q_v R_v}{C_{pd} \frac{R^*}{R_d}}} 
\end{equation}
which is independent of $p$ and therefore we see that $\theta_v$ is an entropic variable for the constant $\kappa$ system. In fact, (\ref{theta-v-expr}) is just another way of writing (\ref{alt-theta-v-expr}).

\paragraph{Dry heat capacities system}
Finally, consider the dry heat capacities system introduced in Section \ref{dry-heat-capacities}. In this case we have
\begin{eqnarray}
         p(\alpha, T, q_d, q_v, q_l, q_i) &=& \frac{R^* T}{\alpha} \\
         p(\alpha, \theta, q_d, q_v, q_l, q_i) &=& \frac{R^*}{\alpha} (\theta)^{\gamma_d} \left( \frac{R^* \theta}{\alpha p_r}\right)^{\frac{R^*}{C_{vd}}} (T_r)^{\frac{R^* - R_d}{C_{vd}}} \\
       p(\alpha, \Pi, q_d, q_v, q_l, q_i) &=& p_r (\frac{\Pi}{C_{pd}})^{\frac{C_{pd}}{R^*}}
\end{eqnarray}
and therefore
\begin{eqnarray}
    T_v(\alpha, p) &=& \frac{p \alpha}{R_d} \\
    \theta_v(\alpha, p) &=&  (\frac{p_r}{p})^{\kappa_d} \frac{p \alpha}{R_d}\\
\Pi_v(\alpha, p) &=& C_{pd} (\frac{p}{p_r})^{\kappa_d}
\end{eqnarray}
Using the expression for $\alpha = \alpha(p,\eta,q_n)$ to obtain $\theta_v(p,\eta,q_n)$ gives
\begin{equation}
    \theta_v(p,\eta,q_n) =  (\frac{p_r}{p})^{\kappa_d} (\frac{p}{R^*})^{\frac{R^*}{C_{pd}}} \frac{R^*}{R_d} T_r \exp^{\frac{\eta - \eta_r}{C_{pd}}} \left( \frac{q_d R_d}{p_{rd}}\right)^{\frac{q_d R_d}{C_{pd}}}  \left( \frac{q_v R_v}{p_{rv}}\right)^{\frac{q_v R_v}{C_{pd}}} 
\end{equation}
and therefore we see that $\theta_v$ is not an entropic variable for the dry heat capacities system.

It is interesting to note that in all three systems, the expression $p \alpha = R^* T = R_d T_v = \kappa_d \theta_v \Pi_v$ holds, which is usually taken as the starting point for definition of $\theta_v$. In fact, all the expressions for $\theta_v$, $T_v$ and $\Pi_v$ in terms of $\alpha$ and $p$ are the same. However, only in the case of the constant $\kappa$ system is $\theta_v$ an entropic variable, seen by expressing it in terms of natural variables. This is another strong illustration of the point that derived thermodynamic expressions (referred to as equations of state in the atmospheric dynamics literature) do not express the full thermodynamics of a system.

%\todo{SHOW THAT $\theta_v$ is not an entropic variable, and also work out for constant $\kappa$ case}

%However, as discussed in SECTION, this is not an entropic variable. Only under the constant $\kappa$ approximation does it become an entropic variable, which can be seen by inserting the expression XXX for $p(\alpha, \eta, q_n)$ (from the constant $\kappa$ approximation) into XXX (which is actually the same in both the constant $\kappa$ and unapproximated cases) to obtain XXX, which is independent of $\alpha$. If we instead insert the expression XXX for $p(\alpha, \eta, q_n$ into XXX, the resulting equation is not independent of $\alpha$.

\section{Commonly Used Thermodynamic Quantities}
\label{common-thermo-quantities}

Following classical equilibrium thermodynamics, a variety of commonly used thermodynamic quantities can be defined using the general thermodynamic potentials from Section \ref{review-equilbrium-thermo}: \textit{heat capacities} at constant volume $C_v$ and pressure $C_p$, \textit{sound speed} (squared) $c_s^2$, \textit{adiabatic temperature gradient} $\Gamma$\footnote{This quantity is not often considered in the literature, but it does appear, for example, in \cite{GayBalmaz2017}.}, \textit{coefficient of thermal expansion} $\alpha_p$, \textit{isochoric thermal pressure coefficient} $\beta_v$, \textit{isothermal compressibility} $\beta_T$ and \textit{isentropic compressibility} $\beta_s$

\begin{eqnarray}
\label{cv-general}
C_v(\alpha,T,q_n) = T \pp{\eta}{T}(\alpha,T,q_n) \quad\quad\quad C_p(p,T,q_n) = T \pp{\eta}{T}(p,T,q_n)\\
c_s^2(\alpha,\eta,q_n) = -\alpha^2 \pp{p}{\alpha}(\alpha,\eta,q_n)\quad\quad\quad \Gamma(p,\eta,q_n) = \pp{T}{p}(p,\eta,q_n) \\
\alpha_p(p,T,q_n) = \frac{1}{\alpha} \pp{\alpha}{T}(p,T,q_n) \quad\quad
\beta_v(\alpha,T,q_n) = \frac{1}{p} \pp{p}{T}(\alpha,T,q_n) \\
\label{beta-general}
\beta_T(p,T,q_n) = -\frac{1}{\alpha} \pp{\alpha}{p}(p,T,q_n) \quad\quad
\beta_s(p,\eta,q_n) = -\frac{1}{\alpha} \pp{\alpha}{p}(p,\eta,q_n)
\end{eqnarray}
along with \textit{adiabatic index} $\gamma = \frac{C_p}{C_v}$.

The quantities (\ref{cv-general}) - (\ref{beta-general}) can be used to rewrite the fundamental thermodynamic identities for $u$ (\ref{U-fund-thermo}) and $h$ (\ref{H-fund-thermo}) in terms of $dT$ instead of $d\eta$ as
\begin{eqnarray}
\label{dU-T}
du &=& C_v dT + p(\beta_v T -1) d\alpha + \sum_n (\mu_n + T \pp{\eta}{q_n}) dq_n \\
\label{dH-T}
dh &=& C_p dT + \alpha(1-\alpha_p T)dp + \sum_n (\mu_n + T \pp{\eta}{q_n}) dq_n
\end{eqnarray}
This requires expressing $d\eta$ in terms of $dT$, $dq_n$, either $d\alpha$ or $dp$, and appropriate partial derivatives of $\eta$ by using $\eta(\alpha,T,q_n)$ or $\eta(p,T,q_n)$, and then using (\ref{cv-general}) - (\ref{beta-general}) to simplify. Note that we have $\pp{\eta}{q_n}(\alpha,T,q_n)$ in (\ref{dU-T}) and $\pp{\eta}{q_n}(p,T,q_n)$ in (\ref{dH-T}).

\subsection{Maxwell Relationships}
Using the equality of mixed partial derivatives, a set of relationships between the derivatives of $p$, $\alpha$, $\eta$ and $T$ can be established:
\begin{eqnarray}
\frac{\partial^2 u}{\partial \alpha \partial \eta} = -\pp{p}{\eta} = \pp{T}{\alpha}\quad\quad\quad\quad
\frac{\partial^2 h}{\partial p \partial \eta} = \pp{\alpha}{\eta} = \pp{T}{p} \\
\frac{\partial^2 g}{\partial p \partial T} = \pp{\alpha}{T} = -\pp{\eta}{p} \quad\quad\quad\quad
\frac{\partial^2 f}{\partial \alpha \partial T} = \pp{p}{T} = \pp{\eta}{\alpha}
\end{eqnarray}

Similarly, for an arbitrary entropic variable $\chi$ a set of relationships between the derivatives of $p$, $\alpha$, $\chi$ and $\lambda$ can be established:
\begin{eqnarray}
\frac{\partial^2 u}{\partial \alpha \partial \chi} = -\pp{p}{\chi} = \pp{\lambda}{\alpha}\quad\quad\quad\quad
\frac{\partial^2 h}{\partial p \partial \chi} = \pp{\alpha}{\chi} = \pp{\lambda}{p}
\end{eqnarray}

These are known as \textit{Maxwell relationships}, and are quite useful in determining the commonly used thermodynamic quantities and relating partial derivatives.

\subsection{\textit{Unapproximated} system}
Now, based on the \textit{unapproximated} thermodynamic potentials from Section \ref{unapprox-moist-air}, the commonly used thermodynamic quantities are
\begin{eqnarray}
C_v(\alpha,T,q_d,q_v,q_l,q_i) &=& T \frac{C_v^*}{T} = C_v^*\\
C_p(p,T,q_d,q_v,q_l,q_i) &=& T \frac{C_p^*}{T} = C_p^*\\
\Gamma(p,\eta,q_d,q_v,q_l,q_i) &=& \frac{R^*}{p C_p^*} T_r \exp (\frac{\eta - \eta_r}{C_p^*}) (\frac{q_d R_d p}{R^* p_{rd}})^{\frac{q_d R_d}{C_p^*}} (\frac{q_v R_v p}{R^* p_{rv}})^{\frac{q_v R_v}{C_p^*}} = \frac{R^* T}{p C_p^*} = \frac{\alpha}{C_p^*}\\
c_s^2(\alpha,\eta,q_d,q_v,q_l,q_i) &=& \frac{R^* C_p^*}{C_v^*} T_r \exp (\frac{\eta - \eta_r}{C_v^*}) (\frac{\alpha}{q_d \alpha_{rd}})^{-\frac{q_d R_d}{C_v^*}} (\frac{\alpha}{q_v \alpha_{rv}})^{-\frac{q_v R_v}{C_v^*}} = \frac{R^* C_p^* T}{C_v^*}\\
\alpha_p(p,T,q_d,q_v,q_l,q_i) &=& \frac{1}{T}\\
\beta_v(\alpha,T,q_d,q_v,q_l,q_i) &=& \frac{1}{T}\\
\beta_T(p,T,q_d,q_v,q_l,q_i) &=& \frac{1}{p}\\
\beta_s(p,\eta,q_d,q_v,q_l,q_i) &=& \frac{C_v^*}{C_p^* p}
\end{eqnarray}
As expected, $C_v = C_v^*$ and $C_p = C_p^*$. Note that these together give $\frac{1}{\alpha^2} c_s^2 C_v \Gamma = \frac{1}{\alpha^2} c_s^2 C_v^* \Gamma = p$. Additionally, we have
\begin{eqnarray}
    du &=& C_v^* dT + C_{vd} (T - T_r) dq_d + C_{vv} (T - T_r) dq_v + C_{l} (T - T_r) dq_l  + C_{i} (T - T_r) dq_i - R_v T_r dq_v + (L_{vr} + L_{fr}) dq_v + L_{fr} dq_l \\
    dh &=& C_p^* dT + C_{pd} (T - T_r) dq_d + C_{pv} (T - T_r) dq_v + C_{l} (T - T_r) dq_l  + C_{i} (T - T_r) dq_i + R_d T_r dq_d + (L_{vr} + L_{fr}) dq_v + L_{fr} dq_l
\end{eqnarray}
along with $C_v + R^* = C_p$.

\subsection{Constant $\kappa$ system}
Based on the constant $\kappa$ thermodynamic potentials from Section \ref{constant-kappa}, the commonly used thermodynamic quantities are
\begin{eqnarray}
C_v(\alpha,T,q_d,q_v,q_l,q_i) &=& \frac{C_{vd} R^*}{R_d}\\
C_p(p,T,q_d,q_v,q_l,q_i) &=& \frac{C_{pd} R^*}{R_d}\\
\Gamma(p,\eta,q_d,q_v,q_l,q_i) &=& \frac{R_d}{p C_{pd}} T_r \exp (\frac{\eta - \eta_r}{C_{pd} \frac{R^*}{R_d}}) (\frac{p}{R^*})^{\frac{R_d}{C_{pd}}} (\frac{q_d R_d}{p_{rd}})^{\frac{q_d R_d}{C_{pd} \frac{R^*}{R_d}}} (\frac{q_v R_v}{p_{rv}})^{\frac{q_v R_v}{C_{pd} \frac{R^*}{R_d}}} =\frac{R_d T}{p C_{pd}} = \frac{R_d \alpha}{R^* C_{pd}}\\
c_s^2(\alpha,\eta,q_d,q_v,q_l,q_i) &=& \frac{C_{pd}}{C_{vd}} \exp (\frac{\eta - \eta_r}{C_{vd} \frac{R^*}{R_d}}) (\alpha)^{-\frac{R_d}{C_{vd}}} (\frac{1}{q_d \alpha_{rd}})^{-\frac{q_d R_d}{C_{vd} \frac{R^*}{R_d}}} (\frac{1}{q_v \alpha_{rv}})^{-\frac{q_v R_v}{C_{vd} \frac{R^*}{R_d}}} = \frac{R^* C_{pd} T}{C_{vd}}\\
\alpha_p(p,T,q_d,q_v,q_l,q_i) &=& \frac{1}{T}\\
\beta_v(\alpha,T,q_d,q_v,q_l,q_i) &=& \frac{1}{T}\\
\beta_T(p,T,q_d,q_v,q_l,q_i) &=& \frac{1}{p}\\
\beta_s(p,\eta,q_d,q_v,q_l,q_i) &=& \frac{C_{vd}}{C_{pd} p}
\end{eqnarray}
As expected, $C_v = \frac{C_{vd} R^*}{R_d}$ and $C_p = \frac{C_{pd} R^*}{R_d}$. Note that these together give $\frac{1}{\alpha^2} c_s^2 C_v \Gamma = \frac{1}{\alpha^2} c_s^2 \frac{C_{vd} R^*}{R_d} \Gamma = p$. Additionally, we have
\begin{eqnarray}
    du &=& \frac{C_{vd} R^*}{R_d} dT + C_{vd} (T - T_r) dq_d + C_{vd} \frac{R_v}{R_d} (T - T_r) dq_v - R_v T_r dq_v + (L_{vr} + L_{fr}) dq_v + L_{fr} dq_l\\
    dh &=& \frac{C_{pd} R^*}{R_d} dT + C_{pd} (T - T_r) dq_d + C_{pd} \frac{R_v}{R_d} (T - T_r) dq_v + R_d T_r dq_d + (L_{vr} + L_{fr}) dq_v + L_{fr} dq_l
\end{eqnarray}
along with $C_v + R^* = C_p$.

\subsection{Dry Heat Capacities system}
Based on the dry heat capacities thermodynamic potentials from Section \ref{dry-heat-capacities}, the commonly used thermodynamic quantities are
\begin{eqnarray}
C_v(\alpha,T,q_d,q_v,q_l,q_i) &=& T \frac{C_{vd}}{T} = C_{vd}\\
C_p(p,T,q_d,q_v,q_l,q_i) &=& T \frac{C_{pd}}{T} = C_{pd}\\
\Gamma(p,\eta,q_d,q_v,q_l,q_i) &=& \frac{R^*}{p C_{pd}} T_r \exp (\frac{\eta - \eta_r}{C_{pd}}) (\frac{q_d R_d p}{R^* p_{rd}})^{\frac{q_d R_d}{C_{pd}}} (\frac{q_v R_v p}{R^* p_{rv}})^{\frac{q_v R_v}{C_{pd}}} = \frac{R^* T}{p C_{pd}} = \frac{\alpha}{C_{pd}}\\
c_s^2(\alpha,\eta,q_d,q_v,q_l,q_i) &=& \frac{R^* C_{pd}}{C_{vd}} T_r \exp (\frac{\eta - \eta_r}{C_{vd}}) (\frac{\alpha}{q_d \alpha_{rd}})^{-\frac{q_d R_d}{C_{vd}}} (\frac{\alpha}{q_v \alpha_{rv}})^{-\frac{q_v R_v}{C_{vd}}} = \frac{R^* C_{pd} T}{C_{vd}}\\
\alpha_p(p,T,q_d,q_v,q_l,q_i) &=& \frac{1}{T}\\
\beta_v(\alpha,T,q_d,q_v,q_l,q_i) &=& \frac{1}{T}\\
\beta_T(p,T,q_d,q_v,q_l,q_i) &=& \frac{1}{p}\\
\beta_s(p,\eta,q_d,q_v,q_l,q_i) &=& \frac{C_{vd}}{C_{pd} p}
\end{eqnarray}
As expected, $C_v = C_{vd}$ and $C_p = C_{pd}$. Note that these together give $\frac{1}{\alpha^2} c_s^2 C_{vd} \Gamma = p$. Additionally, we have
\begin{eqnarray}
    du &=& C_{vd} dT - R_v T_r dq_v + (L_{vr} + L_{fr}) dq_v + L_{fr} dq_l \\
    dh &=& C_{pd} dT + R_d T_r dq_d + (L_{vr} + L_{fr}) dq_v + L_{fr} dq_l
\end{eqnarray}
However, we have $C_v + R^* \neq C_p$, in contrast to unapproximated and the constant $\kappa$ approximation cases.

\section{Latent Heats}
\label{latent-heats}

To determine the latent heats (which are simply the partial enthalpy differences between phases), split $h$ into contributions due to dry air, water vapor, liquid and ice through
\begin{equation}
    h = q_d h_d + q_v h_v + q_l h_l + q_i h_i
\end{equation}
In general, we have $h_k(p, \eta, q_d, q_v, q_l, q_i)$ in this definition, not $h_k(p, \eta, q_k)$. Then the \textit{latent heat of vaporization} $L_v$ and \textit{latent heat of melting} $L_f$ are defined as:
\begin{eqnarray}
    L_v &=& h_v - h_l \\
    L_f &=& h_l - h_i
\end{eqnarray}
These are worked out for the various thermodynamic potentials below, using the form $h = h(T, q_d, q_v, q_l, q_i)$.

\subsection{\textit{Unapproximated} system}
Based on the \textit{unapproximated} thermodynamic potentials from Section \ref{unapprox-moist-air}, the partial enthalpies are
\begin{eqnarray}
h_v &=& C_{pv} (T-T_r) + L_{vr} + L_{fr}\\
h_l &=& C_l (T-T_r) + L_{fr}\\
h_i &=& C_l (T-T_r)
\end{eqnarray}
which gives 
\begin{eqnarray}
    L_v &=& (C_{pv} - C_l) T + L_{vr} - (C_{pv} - C_l) T_r \\
    L_f &=& (C_{l} - C_i) T + L_{fr} - (C_{l} - C_i) T_r
\end{eqnarray}
At $T=T_r$, we have $L_v = L_{vr}$ and $L_f = L_{fr}$, as expected. This fits with the definitions in \cite{Thuburn2017} if we define
\begin{eqnarray}
L_v^0 &=& L_{vr} - (C_{pv} - C_l) T_r\\
L_f^0 &=& L_{fr} - (C_{l} - C_i) T_r
\end{eqnarray}
to finally get
\begin{eqnarray}
    L_v &=& (C_{pv} - C_l) T + L_v^0 \\
    L_f &=& (C_{l} - C_i) T + L_f^0
\end{eqnarray}

\subsection{Constant $\kappa$ system}
Based on the constant $\kappa$ thermodynamic potentials from Section \ref{constant-kappa}, the partial enthalpies are
\begin{eqnarray}
h_v &=& \frac{C_{pd} R_v}{R_d} (T-T_r) + L_{vr} + L_{fr}\\
h_l &=& L_{fr}\\
h_i &=& 0 
\end{eqnarray}
which gives 
\begin{eqnarray}
    L_v &=& \frac{C_{pd} R_v}{R_d} T + L_{vr} - \frac{C_{pd} R_v}{R_d} T_r \\
    L_f &=& L_{fr}
\end{eqnarray}
At $T=T_r$, we have $L_v = L_{vr}$ and $L_f = L_{fr}$, as expected. This fits with the definitions in \cite{Thuburn2017} if we define
\begin{eqnarray}
L_v^0 &=& L_{vr} - \frac{C_{pd} R_v}{R_d} T_r \\
L_f^0 &=& L_{fr}
\end{eqnarray}
to finally get
\begin{eqnarray}
    L_v &=& \frac{C_{pd} R_v}{R_d} T + L_v^0 \\
    L_f &=& L_f^0
\end{eqnarray}
From the perspective of latent heats, the constant $\kappa$ approximation consists of dropping the partial enthalpies associated with liquid/ice (other than the $ L_{fr}$ term in $h_l$) and modifying $C_{pv}$ to $\frac{C_{pd} R_v}{R_d}$. This eliminates the temperature dependence of the latent heat of melting and changes the coefficient of temperature dependence for the latent heat of vaporization.

\subsection{Dry Heat Capacities system}
Based on the dry heat capacities thermodynamic potentials from Section \ref{dry-heat-capacities}, the partial enthalpies are
\begin{eqnarray}
h_v &=& C_{pd} (T-T_r) + L_{vr} + L_{fr}\\
h_l &=& C_{pd} (T-T_r) + L_{fr}\\
h_i &=& C_{pd} (T-T_r)
\end{eqnarray} 
where we have used $\sum_n q_n = 1$, which gives 
\begin{eqnarray}
    L_v &=& L_{vr} \\
    L_f &=& L_{fr}
\end{eqnarray}
This fits with the definitions in \cite{Thuburn2017} if we define
\begin{eqnarray}
L_v^0 &=& L_{vr} \\
L_f^0 &=& L_{fr}
\end{eqnarray}
to finally get
\begin{eqnarray}
    L_v &=& L_v^0 \\
    L_f &=& L_f^0
\end{eqnarray}
As expected, all temperature dependence in the latent heats has disappeared.

\section{Chemical Potentials $\mu_n$ and Generalized Chemical Potentials $\xi_n$}
\label{chemical-potentials}
The expressions for chemical potential $\mu_n$ and generalized chemical potentials $\xi_n$ are quite complicated, and are given here to avoid disrupting the flow of the main text. 

Start by considering the expression of generalized chemical potentials $\xi_n := \pp{X^\prime}{M_n}$ for $X^\prime \in \{U^\prime,H^\prime\}$. By using (\ref{S-entropic}), (\ref{full-to-full-entropic}) and the chain rule, these are given as
\begin{equation}
\label{xi-n-eqn}
    \xi_n  = \mu_n - \lambda \chi + \eta T  + T \pp{\eta}{q_n} - \sum_{n^\prime} T \pp{\eta}{q_{n^\prime}} q_{n^\prime}
\end{equation}
where we have used $\mu_n := \pp{X}{M_n}$.

Next, using (\ref{specific-to-full-potentials}), the chain rule and some algebra, it is not too difficult to obtain an expression for chemical potential $\mu_n$ (which is $\pp{X}{M_n}$ for $X \in \{U,H,G,F\}$) in terms of $\pp{x}{q_n}$ for $x \in \{u,h,g,f\}$:
\begin{equation}
    \mu_n := \pp{X}{M_n} = g + \pp{x}{q_n} - \sum_{n^\prime} q_{n^\prime} \pp{x}{q_{n^\prime}}
\end{equation}
Similarly, using (\ref{full-entropic-to-specific-entropic}) we can obtain an expression for generalized chemical potential $\xi_n$ in terms of $\pp{x^\prime}{q_n}$ for $x \in \{u^\prime,h^\prime\}$
\begin{equation}
        \xi_n := \pp{X^\prime}{M_n} = g_\chi + \pp{x^\prime}{q_n} - \sum_{n^\prime} q_{n^\prime} \pp{x^\prime}{q_{n^\prime}}
\end{equation}
where $g_\chi = u^\prime + p \alpha - \lambda \chi$. Note that
\begin{equation}
    \sum_n q_n \mu_n = g \quad\quad\quad \sum_n q_n \xi_n = g_\chi
\end{equation}

To close these expressions, we need $\pp{x}{q_n}$ for $x \in \{u,h,g,f\}$ and $\pp{x^\prime}{q_n}$ for $x^\prime \in \{u^\prime,h^\prime\}$ in terms of the relevant state variables. These are provided in the following sections.

\begin{remark}
An important point is that not all component specific concentrations are independent, since $\sum_k q_k = 1$, so there are actually only $N-1$ independent concentrations. However, the component masses $M_n$ are independent, and since the chemical potentials are fundamentally defined in terms of $M_n$, this interdependence between the $q_n$'s does not matter, and we can treat all the $q_n$'s as independent when taking derivatives.
\end{remark}

\subsection{\textit{Unapproximated} system}
\label{chemical-potentials-unapprox}
\subsubsection{$\pp{x}{q_n}$ for $x \in \{u,h,g,f\}$}
Based on the \textit{unapproximated} thermodynamic potentials from Section \ref{unapprox-moist-air}, $\pp{x}{q_n}$ for $x \in \{u,h,g,f\}$ are:
\begin{eqnarray}
\pp{u}{q_d}(\alpha,\eta,q_d,q_v,q_l,q_i) &=& A \left[ C_{vd} - \frac{C_{vd}}{C_v^*} \left( \eta - \eta_r - q_d R_d \ln \frac{\alpha}{q_d \alpha_{rd}} - q_v R_v \ln \frac{\alpha}{q_v \alpha_{rv}} \right) + R_d \ln \frac{q_d \alpha_{rd}}{\alpha} + R_d - \eta_{rd} \right] - C_{vd} T_r \\
\pp{u}{q_v}(\alpha,\eta,q_d,q_v,q_l,q_i) &=& A \left[ C_{vv} - \frac{C_{vv}}{C_v^*} \left( \eta - \eta_r - q_d R_d \ln \frac{\alpha}{q_d \alpha_{rd}} - q_v R_v \ln \frac{\alpha}{q_v \alpha_{rv}} \right) + R_v \ln \frac{q_v \alpha_{rv}}{\alpha} + R_v - \eta_{rv}\right] - C_{vv} T_r + L_{vr} + L_{fr} - R_v T_r\\
\pp{u}{q_l}(\alpha,\eta,q_d,q_v,q_l,q_i) &=& A \left[ C_l - \frac{C_l}{C_v^*} \left( \eta - \eta_r - q_d R_d \ln \frac{\alpha}{q_d \alpha_{rd}} - q_v R_v \ln \frac{\alpha}{q_v \alpha_{rv}} \right) - \eta_{rl} \right] - C_l T_r + L_{fr}\\
\pp{u}{q_i}(\alpha,\eta,q_d,q_v,q_l,q_i) &=& A \left[ C_i - \frac{C_i}{C_v^*} \left( \eta - \eta_r - q_d R_d \ln \frac{\alpha}{q_d \alpha_{rd}} - q_v R_v \ln \frac{\alpha}{q_v \alpha_{rv}} \right) - \eta_{ri} \right] - C_i T_r
\end{eqnarray}
with $A = T_r \exp (\frac{\eta - \eta_r}{C_v^*}) (\frac{\alpha}{q_d \alpha_{rd}})^{-\frac{q_d R_d}{C_v^*}} (\frac{\alpha}{q_v \alpha_{rv}})^{-\frac{q_v R_v}{C_v^*}} = T(\alpha, \eta, q_n)$.

\begin{eqnarray}
\pp{h}{q_d}(p,\eta,q_d,q_v,q_l,q_i) &=& A \left[ C_{pd} - \frac{C_{pd}}{C_p^*} \left( \eta - \eta_r + q_d R_d \ln \frac{q_d R_d p}{R^* p_{rd}} + q_v R_v \ln \frac{q_v R_v p}{R^* p_{rv}} \right) + R_d \ln \frac{q_d R_d p}{R^* p_{rd}} - \eta_{rd} \right] - C_{pd} T_r + R_d T_r\\
\pp{h}{q_v}(p,\eta,q_d,q_v,q_l,q_i) &=& A \left[ C_{pv} - \frac{C_{pv}}{C_p^*} \left( \eta - \eta_r + q_d R_d \ln \frac{q_d R_d p}{R^* p_{rd}} + q_v R_v \ln \frac{q_v R_v p}{R^* p_{rv}} \right) + R_v \ln \frac{q_v R_v p}{R^* p_{rv}} - \eta_{rv} \right] - C_{pv} T_r + L_{vr} + L_{fr}\\
\pp{h}{q_l}(p,\eta,q_d,q_v,q_l,q_i) &=& A  \left[ C_l - \frac{C_l}{C_p^*} \left( \eta - \eta_r + q_d R_d \ln \frac{q_d R_d p}{R^* p_{rd}} + q_v R_v \ln \frac{q_v R_v p}{R^* p_{rv}} \right) - \eta_{rl} \right] - C_l T_r + L_{fr}\\
\pp{h}{q_i}(p,\eta,q_d,q_v,q_l,q_i) &=& A  \left[ C_i - \frac{C_i}{C_p^*} \left( \eta - \eta_r + q_d R_d \ln \frac{q_d R_d p}{R^* p_{rd}} + q_v R_v \ln \frac{q_v R_v p}{R^* p_{rv}} \right) - \eta_{ri} \right] - C_i T_r 
\end{eqnarray}
with $A = T_r \exp (\frac{\eta - \eta_r}{C_p^*}) (\frac{q_d R_d p}{R^* p_{rd}})^{\frac{q_d R_d}{C_p^*}} (\frac{q_v R_v p}{R^* p_{rv}})^{\frac{q_v R_v}{C_p^*}} = T(p, \eta, q_n)$.

\begin{eqnarray}
\pp{g}{q_d}(p,T,q_d,q_v,q_l,q_i) &=& T (C_{pd} - C_{pd} \ln \frac{T}{T_r} + R_d \ln \frac{q_d R_d p}{R^* p_{rd}} - \eta_{rd}) - C_{pd} T_r + R_d T_r\\
\pp{g}{q_v}(p,T,q_d,q_v,q_l,q_i) &=& T (C_{pv} - C_{pv} \ln \frac{T}{T_r} + R_v \ln \frac{q_v R_v p}{R^* p_{rv}} - \eta_{rv}) - C_{pv} T_r + L_{vr} + L_{fr}\\
\pp{g}{q_l}(p,T,q_d,q_v,q_l,q_i) &=& T (C_l - C_l \ln \frac{T}{T_r} - \eta_{rl}) - C_l T_r + L_{fr}\\
\pp{g}{q_i}(p,T,q_d,q_v,q_l,q_i) &=& T (C_i - C_i \ln \frac{T}{T_r} - \eta_{ri}) - C_i T_r
\end{eqnarray}

\begin{eqnarray}
\pp{f}{q_d}(\alpha,T,q_d,q_v,q_l,q_i) &=& T (C_{vd} - C_{vd} \ln \frac{T}{T_r} + R_d \ln \frac{q_d \alpha_{rd}}{\alpha} + R_d - \eta_{rd}) - C_{vd} T_r\\
\pp{f}{q_v}(\alpha,T,q_d,q_v,q_l,q_i) &=& T (C_{vv} - C_{vv} \ln \frac{T}{T_r} + R_v \ln \frac{q_v \alpha_{rv}}{\alpha} + R_v - \eta_{rv}) - C_{vv} T_r + L_{vr} + L_{fr} - R_v T_r\\
\pp{f}{q_l}(\alpha,T,q_d,q_v,q_l,q_i) &=& T(C_l - C_l \ln \frac{T}{T_r} - \eta_{rl}) - C_l T_r + L_{fr}\\
\pp{f}{q_i}(\alpha,T,q_d,q_v,q_l,q_i) &=& T(C_i - C_i \ln \frac{T}{T_r} - \eta_{ri}) - C_i T_r
\end{eqnarray}

\subsubsection{$\pp{x^\prime}{q_n}$ for $x^\prime \in \{u^\prime,h^\prime\}$ for Potential Temperature $\theta$}
Based on the thermodynamic potentials from Section \ref{unapprox-moist-air}, $\pp{x^\prime}{q_n}$ for $x^\prime \in \{u^\prime,h^\prime\}$ for potential temperature $\theta$ are:

\begin{eqnarray}
\pp{u^\prime}{q_d}(\alpha,\theta,q_d,q_v,q_l,q_i) &=& A \left[ C_{vd} + \ln (\theta) (C_{pd} - \frac{C_{vd} C_p^*}{C_v^*}) + \ln (\frac{R^*}{\alpha p_r}) (R_d - \frac{C_{vd} R^*}{C_v^*})\right] - C_{vd} T_r \\
\pp{u^\prime}{q_v}(\alpha,\theta,q_d,q_v,q_l,q_i) &=& A \left[ C_{vv} + \ln (\theta) (C_{pv} - \frac{C_{vv} C_p^*}{C_v^*}) + \ln (\frac{R^*}{\alpha p_r}) (R_v - \frac{C_{vv} R^*}{C_v^*})\right] - C_{vv} T_r + L_{vr} + L_{fr} - R_v T_r \\
\pp{u^\prime}{q_l}(\alpha,\theta,q_d,q_v,q_l,q_i) &=& A \left[ C_{l} + \ln (\theta) (C_{l} - \frac{C_{l} C_p^*}{C_v^*}) + \ln (\frac{R^*}{\alpha p_r}) (-\frac{C_{l} R^*}{C_v^*})\right] - C_{l} T_r + L_{fr}\\
\pp{u^\prime}{q_i}(\alpha,\theta,q_d,q_v,q_l,q_i) &=& A \left[ C_{i} + \ln (\theta) (C_{i} - \frac{C_{i} C_p^*}{C_v^*}) + \ln (\frac{R^*}{\alpha p_r}) (-\frac{C_{i} R^*}{C_v^*})\right] - C_{i} T_r
\end{eqnarray}
with $A =(\theta)^{\gamma^*} (\frac{R^*}{\alpha p_r})^{\delta^*} = T(\alpha, \theta, q_n)$.

\begin{eqnarray}
\pp{h^\prime}{q_d}(p,\theta,q_d,q_v,q_l,q_i) &=& \theta (\frac{p}{p_r})^{\kappa^*} \left[ C_{pd} + \ln (\frac{p}{p_r}) (R_d - \frac{R^* C_{pd}}{C_p^*})\right] - C_{vd} T_r \\
\pp{h^\prime}{q_v}(p,\theta,q_d,q_v,q_l,q_i) &=& \theta (\frac{p}{p_r})^{\kappa^*} \left[ C_{pv} - \ln (\frac{p}{p_r}) (R_v - \frac{R^* C_{pv}}{C_p^*})\right] - C_{vv} T_r + L_{vr} + L_{fr} - R_v T_r \\
\pp{h^\prime}{q_l}(p,\theta,q_d,q_v,q_l,q_i) &=& \theta (\frac{p}{p_r})^{\kappa^*} \left[ C_{l} + \ln (\frac{p}{p_r}) \frac{R^* C_{l}}{C_p^*})\right] - C_{l} T_r + L_{fr}\\
\pp{h^\prime}{q_i}(p,\theta,q_d,q_v,q_l,q_i) &=& \theta (\frac{p}{p_r})^{\kappa^*} \left[ C_{i} + \ln (\frac{p}{p_r}) \frac{R^* C_{i}}{C_p^*})\right] - C_{i} T_r 
\end{eqnarray}

Note that $\theta (\frac{p}{p_r})^{\kappa^*} = T(p,\theta,q_n)$.

\subsection{Constant $\kappa$ system}
\label{chemical-potentials-const-kappa}

\subsubsection{$\pp{x}{q_n}$ for $x \in \{u,h,g,f\}$}
Based on the constant $\kappa$ thermodynamic potentials from Section \ref{constant-kappa}, $\pp{x}{q_n}$ for $x \in \{u,h,g,f\}$ are:

\begin{eqnarray}
\pp{u}{q_d}(\alpha,\eta,q_d,q_v,q_l,q_i) &=& A \left[ C_{vd} - \frac{R_d}{R^*} \left( \eta - \eta_r - q_d R_d \ln \frac{\alpha}{q_d \alpha_{rd}} - q_v R_v \ln \frac{\alpha}{q_v \alpha_{rv}} \right) + R_d \ln \frac{q_d \alpha_{rd}}{\alpha} + R_d - \eta_{rd} \right] - C_{vd} T_r \\
\pp{u}{q_v}(\alpha,\eta,q_d,q_v,q_l,q_i) &=& A \left[ \frac{C_{vd} R_v}{R_d} - \frac{R_v}{R^*} \left( \eta - \eta_r - q_d R_d \ln \frac{\alpha}{q_d \alpha_{rd}} - q_v R_v \ln \frac{\alpha}{q_v \alpha_{rv}} \right) + R_v \ln \frac{q_v \alpha_{rv}}{\alpha} + R_v - \eta_{rv}\right] - \frac{C_{vd} R_v}{R_d} T_r + L_{vr} + L_{fr} - R_v T_r\\
\pp{u}{q_l}(\alpha,\eta,q_d,q_v,q_l,q_i) &=& - A \eta_{rl}  + L_{fr}\\
\pp{u}{q_i}(\alpha,\eta,q_d,q_v,q_l,q_i) &=& - A \eta_{ri}
\end{eqnarray}
with $A = T_r \exp (\frac{\eta - \eta_r}{\frac{C_{vd} R^*}{R_d}}) (\frac{\alpha}{q_d \alpha_{rd}})^{-\frac{q_d R_d}{\frac{C_{vd} R^*}{R_d}}} (\frac{\alpha}{q_v \alpha_{rv}})^{-\frac{q_v R_v}{\frac{C_{vd} R^*}{R_d}}} = T(\alpha,\eta,q_n)$.

%\frac{C_{vd} R^*}{R_d}
%\frac{C_{pd} R^*}{R_d}

\begin{eqnarray}
\pp{h}{q_d}(p,\eta,q_d,q_v,q_l,q_i) &=& A \left[ C_{pd} - \frac{R_d}{R^*} \left( \eta - \eta_r + q_d R_d \ln \frac{q_d R_d p}{R^* p_{rd}} + q_v R_v \ln \frac{q_v R_v p}{R^* p_{rv}} \right) + R_d \ln \frac{q_d R_d p}{R^* p_{rd}} - \eta_{rd} \right] - C_{pd} T_r + R_d T_r\\
\pp{h}{q_v}(p,\eta,q_d,q_v,q_l,q_i) &=& A \left[ \frac{C_{pd} R_v}{R_d} - \frac{R_v}{R^*} \left( \eta - \eta_r + q_d R_d \ln \frac{q_d R_d p}{R^* p_{rd}} + q_v R_v \ln \frac{q_v R_v p}{R^* p_{rv}} \right) + R_v \ln \frac{q_v R_v p}{R^* p_{rv}} - \eta_{rv} \right] - \frac{C_{pd} R_v}{R_d} T_r + L_{vr} + L_{fr}\\
\pp{h}{q_l}(p,\eta,q_d,q_v,q_l,q_i) &=& - A \eta_{rl}  + L_{fr}\\
\pp{h}{q_i}(p,\eta,q_d,q_v,q_l,q_i) &=& - A \eta_{ri}
\end{eqnarray}
with $A = T_r \exp (\frac{\eta - \eta_r}{\frac{C_{pd} R^*}{R_d}}) (\frac{q_d R_d p}{R^* p_{rd}})^{\frac{q_d R_d}{\frac{C_{pd} R^*}{R_d}}} (\frac{q_v R_v p}{R^* p_{rv}})^{\frac{q_v R_v}{\frac{C_{pd} R^*}{R_d}}} = T(p,\eta,q_n)$.

\begin{eqnarray}
\pp{g}{q_d}(p,T,q_d,q_v,q_l,q_i) &=& T (C_{pd} - C_{pd} \ln \frac{T}{T_r} + R_d \ln \frac{q_d R_d p}{R^* p_{rd}} - \eta_{rd}) - C_{pd} T_r + R_d T_r\\
\pp{g}{q_v}(p,T,q_d,q_v,q_l,q_i) &=& T (\frac{C_{pd} R_v}{R_d} - \frac{C_{pd} R_v}{R_d} \ln \frac{T}{T_r} + R_v \ln \frac{q_v R_v p}{R^* p_{rv}} - \eta_{rv}) - \frac{C_{pd} R_v}{R_d} T_r + L_{vr} + L_{fr}\\
\pp{g}{q_l}(p,T,q_d,q_v,q_l,q_i) &=& - T \eta_{rl} + L_{fr}\\
\pp{g}{q_i}(p,T,q_d,q_v,q_l,q_i) &=& - T \eta_{ri}
\end{eqnarray}

\begin{eqnarray}
\pp{f}{q_d}(\alpha,T,q_d,q_v,q_l,q_i) &=& T (C_{vd} - C_{vd} \ln \frac{T}{T_r} + R_d \ln \frac{q_d \alpha_{rd}}{\alpha} - \eta_{rd}) - C_{vd} T_r\\
\pp{f}{q_v}(\alpha,T,q_d,q_v,q_l,q_i) &=& T (\frac{C_{vd} R_v}{R_d} - \frac{C_{vd} R_v}{R_d} \ln \frac{T}{T_r} + R_v \ln \frac{q_v \alpha_{rv}}{\alpha} - \eta_{rv}) - \frac{C_{vd} R_v}{R_d} T_r + L_{vr} + L_{fr} - R_v T_r\\
\pp{f}{q_l}(\alpha,T,q_d,q_v,q_l,q_i) &=& - T \eta_{rl} + L_{fr}\\
\pp{f}{q_i}(\alpha,T,q_d,q_v,q_l,q_i) &=& - T \eta_{ri}
\end{eqnarray}

\subsubsection{$\pp{x^\prime}{q_n}$ for $x^\prime \in \{u^\prime,h^\prime\}$ for Potential Temperature $\theta$}
Based on the thermodynamic potentials from Section \ref{constant-kappa}, $\pp{x^\prime}{q_n}$ for $x^\prime \in \{u^\prime,h^\prime\}$ for potential temperature $\theta$ are:
 
 \begin{eqnarray}
\pp{u^\prime}{q_d}(\alpha,\theta,q_d,q_v,q_l,q_i) &=& A \frac{C_{pd}}{R_d} R_d - C_{vd} T_r \\
\pp{u^\prime}{q_v}(\alpha,\theta,q_d,q_v,q_l,q_i) &=& A \frac{C_{pd}}{R_d} R_v - C_{vd} \frac{R_v}{R_d} T_r + L_{vr} + L_{fr} - R_v T_r \\
\pp{u^\prime}{q_l}(\alpha,\theta,q_d,q_v,q_l,q_i) &=& L_{fr}\\
\pp{u^\prime}{q_i}(\alpha,\theta,q_d,q_v,q_l,q_i) &=& 0
\end{eqnarray}
with $A = (\theta)^{\gamma_d} \left( \frac{R^*}{\alpha p_r}\right)^{\delta_d} = T (\alpha, \theta, q_n)$.

\begin{eqnarray}
\pp{h^\prime}{q_d}(p,\theta,q_d,q_v,q_l,q_i) &=& A \frac{C_{pd}}{R_d} R_d - C_{vd} T_r \\
\pp{h^\prime}{q_v}(p,\theta,q_d,q_v,q_l,q_i) &=& A \frac{C_{pd}}{R_d} R_v - C_{pd} \frac{R_v}{R_d} T_r + L_{vr} + L_{fr} \\
\pp{h^\prime}{q_l}(p,\theta,q_d,q_v,q_l,q_i) &=& L_{fr}\\
\pp{h^\prime}{q_i}(p,\theta,q_d,q_v,q_l,q_i) &=& 0
\end{eqnarray}
with $A = \theta \left( \frac{p}{p_r}\right)^{\kappa_d} = T (p, \theta, q_n)$.

\subsubsection{$\pp{x^\prime}{q_n}$ for $x^\prime \in \{u^\prime,h^\prime\}$ for Virtual Potential Temperature $\theta_v$}

Based on the thermodynamic potentials from Section \ref{constant-kappa}, $\pp{x^\prime}{q_n}$ for $x^\prime \in \{u^\prime,h^\prime\}$ for virtual potential temperature $\theta_v$ are:

\begin{eqnarray}
\pp{u^\prime}{q_d}(\alpha,\theta_v,q_d,q_v,q_l,q_i) &=& - C_{vd} T_r \\
\pp{u^\prime}{q_v}(\alpha,\theta_v,q_d,q_v,q_l,q_i) &=& - C_{vd} \frac{R_v}{R_d} T_r + L_{vr} + L_{fr} - R_v T_r \\
\pp{u^\prime}{q_l}(\alpha,\theta_v,q_d,q_v,q_l,q_i) &=& L_{fr} \\
\pp{u^\prime}{q_i}(\alpha,\theta_v,q_d,q_v,q_l,q_i) &=& 0
\end{eqnarray}

\begin{eqnarray}
\pp{h^\prime}{q_d}(p,\theta_v,q_d,q_v,q_l,q_i) &=& - C_{pd} T_r + R_d T_r \\
\pp{h^\prime}{q_v}(p,\theta_v,q_d,q_v,q_l,q_i) &=& - C_{pd} \frac{R_v}{R_d} T_r + L_{vr} + L_{fr} \\
\pp{h^\prime}{q_l}(p,\theta_v,q_d,q_v,q_l,q_i) &=& L_{fr} \\
\pp{h^\prime}{q_i}(p,\theta_v,q_d,q_v,q_l,q_i) &=& 0
\end{eqnarray}

It is key to note that these are all constants, in contrast to the very complicated expressions found for $\pp{X}{q_n}$ and those found for the other systems.

\subsection{Dry Heat Capacities system}
\label{chemical-potentials-dry-heat}

\subsubsection{$\pp{x}{q_n}$ for $x \in \{u,h,g,f\}$}
Based on the dry heat capacities thermodynamic potentials from Section \ref{dry-heat-capacities}, $\pp{x}{q_n}$ for $x \in \{u,h,g,f\}$ are:

\begin{eqnarray}
\pp{u}{q_d}(\alpha,\eta,q_d,q_v,q_l,q_i) &=& A \left[ R_d - R_d \ln \frac{\alpha}{q_d \alpha_{rd}} - \eta_{rd} \right] \\
\pp{u}{q_v}(\alpha,\eta,q_d,q_v,q_l,q_i) &=& A \left[ R_v - R_v \ln \frac{\alpha}{q_v \alpha_{rv}} - \eta_{rv} \right] + L_{vr} + L_{fr} - R_v T_r\\
\pp{u}{q_l}(\alpha,\eta,q_d,q_v,q_l,q_i) &=& - \eta_{rl} X + L_{fr}\\
\pp{u}{q_i}(\alpha,\eta,q_d,q_v,q_l,q_i) &=& - \eta_{ri} X 
\end{eqnarray}
with $A = T_r \exp (\frac{\eta - \eta_r}{C_{vd}}) (\frac{\alpha}{q_d \alpha_{rd}})^{-\frac{q_d R_d}{C_{vd}}} (\frac{\alpha}{q_v \alpha_{rv}})^{-\frac{q_v R_v}{C_{vd}}} = T (\alpha, \eta, q_n)$.

\begin{eqnarray}
\pp{h}{q_d}(p,\eta,q_d,q_v,q_l,q_i) &=& A \left[ \frac{R_d}{C_{pd}} \ln \frac{q_d R_d p}{R^* p_{rd}} - \eta_{rd} \right] + R_d T_r\\
\pp{h}{q_v}(p,\eta,q_d,q_v,q_l,q_i) &=& A \left[ \frac{R_v}{C_{pd}} \ln \frac{q_v R_v p}{R^* p_{rv}} - \eta_{rv} \right] + L_{vr} + L_{fr}\\
\pp{h}{q_l}(p,\eta,q_d,q_v,q_l,q_i) &=& - \eta_{rl} X + L_{fr}\\
\pp{h}{q_i}(p,\eta,q_d,q_v,q_l,q_i) &=& - \eta_{ri} X 
\end{eqnarray}
with $A = T_r \exp (\frac{\eta - \eta_r}{C_{pd}}) (\frac{q_d R_d p}{R^* p_{rd}})^{\frac{q_d R_d}{C_{pd}}} (\frac{q_v R_v p}{R^* p_{rv}})^{\frac{q_v R_v}{C_{pd}}} = T (p, \eta, q_n)$.

\begin{eqnarray}
\pp{g}{q_d}(p,T,q_d,q_v,q_l,q_i) &=& T (R_d \ln \frac{q_d R_d p}{R^* p_{rd}} - \eta_{rd}) + R_d T_r\\
\pp{g}{q_v}(p,T,q_d,q_v,q_l,q_i) &=& T (R_v \ln \frac{q_v R_v p}{R^* p_{rv}} - \eta_{rv}) + L_{vr} + L_{fr}\\
\pp{g}{q_l}(p,T,q_d,q_v,q_l,q_i) &=& -T \eta_{rl} + L_{fr}\\
\pp{g}{q_i}(p,T,q_d,q_v,q_l,q_i) &=& -T \eta_{ri}
\end{eqnarray}

\begin{eqnarray}
\pp{f}{q_d}(\alpha,T,q_d,q_v,q_l,q_i) &=& T (R_d + R_d \ln \frac{q_d \alpha_{rd}}{\alpha} - \eta_{rd})\\
\pp{f}{q_v}(\alpha,T,q_d,q_v,q_l,q_i) &=& T (R_v + R_v \ln \frac{q_v \alpha_{rv}}{\alpha} - \eta_{rv}) + L_{vr} + L_{f} - R_v T_r\\
\pp{f}{q_l}(\alpha,T,q_d,q_v,q_l,q_i) &=& -T \eta_{rl} + L_{fr}\\
\pp{f}{q_i}(\alpha,T,q_d,q_v,q_l,q_i) &=& -T \eta_{ri}
\end{eqnarray}

\subsubsection{$\pp{x^\prime}{q_n}$ for $x^\prime \in \{u^\prime,h^\prime\}$ for Potential Temperature $\theta$}
Based on the thermodynamic potentials from Section \ref{dry-heat-capacities}, $\pp{x^\prime}{q_n}$ for $x^\prime \in \{u^\prime,h^\prime\}$ for potential temperature $\theta$ are:

\begin{eqnarray}
\pp{u^\prime}{q_d}(\alpha,\theta,q_d,q_v,q_l,q_i) &=& A \left[ \ln(\frac{R^*}{\alpha p_r}) + 1 + \ln T_r\right] R_d \\
\pp{u^\prime}{q_v}(\alpha,\theta,q_d,q_v,q_l,q_i) &=& A \left[ \ln(\frac{R^*}{\alpha p_r}) + 1 + \ln T_r\right] R_v + L_{vr} + L_{fr} - R_v T_r \\
\pp{u^\prime}{q_l}(\alpha,\theta,q_d,q_v,q_l,q_i) &=& L_{fr}\\
\pp{u^\prime}{q_i}(\alpha,\theta,q_d,q_v,q_l,q_i) &=& 0
\end{eqnarray}
with $A =(\theta)^{\frac{C_{pd}}{C_{vd}}} (\frac{R^*}{\alpha p_r})^{\frac{R^*}{C_{vd}}} (T_r)^{\frac{R^* - R_d}{C_{vd}}} = T(\alpha, \theta, q_n)$.

\begin{eqnarray}
\pp{h^\prime}{q_d}(p,\theta,q_d,q_v,q_l,q_i) &=& A \ln (\frac{p}{p_r}) R_d + R_d T_r \\
\pp{h^\prime}{q_v}(p,\theta,q_d,q_v,q_l,q_i) &=& A \ln (\frac{p}{p_r}) R_v - C_{vv} T_r + L_{vr} + L_{fr} \\
\pp{h^\prime}{q_l}(p,\theta,q_d,q_v,q_l,q_i) &=& L_{fr}\\
\pp{h^\prime}{q_i}(p,\theta,q_d,q_v,q_l,q_i) &=& 0
\end{eqnarray}
with $A =\theta (\frac{p}{p_r})^{\frac{R^*}{C_{pd}}} = T(p, \theta, q_n) $.

\end{document}